\documentclass[prd,aps,superscriptaddress,showpacs,amssymb,
amsmath,twocolumn,floatfix]{revtex4-1}
\usepackage{graphicx,dcolumn,amsmath,hyperref}

\usepackage[usenames]{color}

\graphicspath{{./}}

\newcommand{\vegas}{\textsc{vegas}}

\renewcommand{\vec}[1]{\boldsymbol{#1}}
\newcommand{\order}{\mathcal{O}}
\newcommand{\con}{\mathrm{con}}
\newcommand{\lat}{\mathrm{lat}}
\newcommand{\tree}{\mathrm{tree}}
\newcommand{\asq}{ASQ}
\newcommand{\QCD}{\mathrm{QCD}}
\newcommand{\asqtad}{ASQTad}
\newcommand{\hisq}{HISQ}
\newcommand{\spinup}{\uparrow}
\newcommand{\spindown}{\downarrow}
\newcommand{\pslash}{{p\negthickspace\negthinspace/\thickspace}}
\newcommand{\kslash}{{k\negthickspace\negthinspace/\thickspace}}
\newcommand{\dd}[1]{\frac{d^4 {#1}}{(2\pi)^4}}
\newcommand{\ocirc}[1]{{\overset{\circ}{#1}}} 
\newcommand{\MeV}{\operatorname{MeV}}
\newcommand{\GeV}{\operatorname{GeV}}
\newcommand{\red}[1]{{\color{red}{#1}}}
\newcommand{\blue}[1]{{\color{blue}{#1}}}
\newcommand{\green}[1]{{\color{green}{#1}}}
\newcommand{\dgamma}{\hat{\gamma}}
\newcommand{\res}{\mathrm{res}}

\newcommand{\cambridge}{Department of Applied Mathematics and
Theoretical Physics, University of Cambridge, Centre for Mathematical
Sciences, Cambridge CB3 0WA, United Kingdom}

\newcommand{\edinburgh}{SUPA, School of Physics and Astronomy,
University of Edinburgh, JCMB, King's Buildings, Mayfield Road, Edinburgh
EH9 3JZ, United Kingdom}

\newcommand{\metoffice}{Met${}\,{}$Office, Fitzroy Road, Exeter EX1 3PB, United Kingdom}
\newcommand{\ERI}{Cray Exascale Research Initiative, JCMB, King's Buildings, Mayfield Road, Edinburgh EH9 3JZ, United Kingdom}

\newcommand{\rel}{\operatorname{rel}}

\begin{document}

\title{Renormalization of heavy-light currents in moving NRQCD}

\author{E.H. \surname{M\"{u}ller}}
\altaffiliation[Current address: ]{\metoffice}
\email{eike.mueller@metoffice.gov.uk}
\affiliation{\edinburgh}

\author{A. \surname{Hart}}
\altaffiliation[Current address: ]{\ERI}
\affiliation{\edinburgh}

\author{R.R. \surname{Horgan}} 
\affiliation{\cambridge}

\date{February 8, 2011}

\collaboration{HPQCD Collaboration}
\noaffiliation

\preprint{Edinburgh 2010/33}

\begin{abstract}
Heavy-light decays such as $B\rightarrow\pi\ell\nu$, $B\rightarrow
K^{*}\gamma$ and $B\rightarrow K^{(*)}\ell\ell$ can be used to
constrain the parameters of the Standard Model and in indirect
searches for new physics. While the precision of experimental results
has improved over the last years this has still to be matched by
equally precise theoretical predictions. The calculation of
heavy-light form factors is currently carried out in lattice QCD.
Due to its small Compton wavelength we discretize the heavy quark in an effective
non-relativistic theory. By formulating the theory in a moving frame of
reference discretization errors in the final state are reduced at
large recoil. Over the last years the formalism has been improved and
tested extensively. Systematic uncertainties are reduced by
renormalizing the m(oving)NRQCD action and heavy-light decay
operators. The theory differs from QCD only for large loop momenta at
the order of the lattice cutoff and the calculation can be carried out
in perturbation theory as an expansion in the strong coupling
constant. In this paper we calculate the one loop corrections to the
heavy-light vector and tensor operator. Due to the complexity of the
action the generation of lattice Feynman rules is automated and loop
integrals are solved by the adaptive Monte Carlo integrator \vegas. We
discuss the infrared and ultraviolet divergences in the loop integrals
both in the continuum and on the lattice. The light quarks are
discretized in the \asqtad\ and highly improved staggered quark
(\hisq) action; the formalism is easily extended to other quark actions.
\end{abstract}

\maketitle

%
%
%
%
%
\section{Introduction}
\label{sec_intro}

Decays of mesons containing heavy quarks provide an excellent 
laboratory for studying the heavy flavor sector of the Standard 
Model. For inclusive decays techniques such as quark-hadron duality 
and the operator product expansion have been used to predict decay 
amplitudes and spectra to high precision (see, for example Refs. 
\cite{Misiak:2006zs,Gardi:2007jx}). Exclusive decays have a well 
defined hadronic final state and are easier to measure but
obtaining precise theoretical predictions for these processes is 
more challenging since the quarks in the final state are bound inside a 
hadron. Nevertheless improving on these predictions is crucial to 
check results from inclusive measurements and overconstrain the 
parameters of the Standard Model to uncover the effects of putative new physics.

The decay \mbox{$B\rightarrow \pi\ell\nu$} can, for instance,
be used to constrain the magnitude of $V_{ub}$, one of the least known
Cabibbo--Kobayashi--Maskawa (CKM) mixing matrix elements. In addition, rare
decays like $B\rightarrow K^*\gamma$ or $B\rightarrow
K^{(*)}\ell\ell$ are loop-suppressed in the Standard Model and
expected to be sensitive to the presence of new physics
\cite{Aliev:1999gp,Aliev:1999re,Neubert:2002ku}.

Precise calculations of hadronic matrix elements are needed to match
experimental precision (which is still to improve further with new
results from LHCb and the planned SuperB factory). Lattice QCD
provides a model-independent framework for these calculations. It is,
however, difficult to discretize the heavy valence quarks directly on
lattices that are currently available since their Compton wavelength is comparable
to the lattice spacing, $a$. To overcome this problem we use the 
effective lattice field theory NRQCD  
\cite{Lepage:1987gg,Thacker:1990bm,Davies:1991py,Lepage:1992tx},
which describes QCD in the non-relativistic limit where the heavy-quark
velocity is much less than unity. NRQCD is useful for studying mesons
with heavy-quark constituents such as $J/\psi$, $\Upsilon$, $D$ and $B$.
The heavy-light vector form factor has been calculated in lattice NRQCD and
combined with experimental data to extract $|V_{ub}|$
\cite{Dalgic:2006dt}.
The hadronic form factors are functions of the squared momentum 
transfer $q^2=(p_B-p_F)^2$, where $p_B$ and $p_F$ are the momenta of 
the decaying $B$ meson and the hadronic final state. Current lattice 
calculations using NRQCD work well only for large $q^2$, partially owing  
to large discretization errors in the hadronic final state. The 
radiative decay $B\rightarrow K^*\gamma$ has $q^2 = 0$, however, and 
most experimental data for $B\rightarrow \pi \ell\nu$ comes from the 
small $q^2$ region. The large extrapolation to small $q^2$ is a sizable 
source of systematic error in analyzing such decays and it is thus 
desirable to extend the range of accessible $q^2$ in a lattice 
calculation. In our approach, this is achieved by formulating the theory in a
moving reference frame
\cite{Sloan:1997fc}.
The frame velocity is chosen to reduce the three-momentum of the final
hadron state at small $q^2$, and hence also suppress the associated
lattice artifacts. This approach is known as moving NRQCD (mNRQCD)
\cite{Foley:2002qv,Foley:2004rp,Foley:2005fx,Horgan:2009ti}.
The momentum of the heavy quark $p=mu+p_{\res}$ 
is split into a contribution $mu$ associated with the frame 
velocity $u=(\gamma,\gamma\vec{v})$ and a residual
momentum $p_{\mathrm{res}}$, which is of the order of the hadronic
scale $\Lambda_{\mathrm{QCD}}$. The first contribution is treated
exactly so that corrections can be expanded in powers of
$p_{\mathrm{res}}/m$. Discretization errors, which scale with some
power of $ap_{\mathrm{res}}$, can be removed systematically by
improving the action. 

Although we focus here on mNRQCD, we note that similar calculations
are possible using other descriptions of moving heavy quarks
\cite{Mandula:1991ds,Hashimoto:1995in,Boyle:2003ui}.
A comparison of predictions from these different approaches would
provide a valuable test of our understanding of the systematic errors
in all these methods.

Hadronic matrix elements of heavy-light currents have been calculated
using lattice mNRQCD by the HPQCD Collaboration
\cite{Meinel:2008th,Liu:2009dj}. 
The NRQCD effective theory is obtained from QCD by integrating out the effects of
physics on energy scales of order $m$ or larger; on the lattice this upper energy scale
is determined by the lattice spacing, $a$. Operators in QCD are written as
a formal expansion of operators defined in terms of the effective non-relativistic fields
of NRQCD. The coefficients of the NRQCD operators in the expansion are determined as
power series in the strong coupling constant $\alpha_s(q^*/a)$ which is defined in an appropriate
renormalization scheme and where $q^*$ is a dimension parameter of order unity. The NRQCD operators
are ordered in powers of $1/m$. For quarkonium matrix elements this gives a series in $\alpha_s$ and
the quark relative velocity $v_{\rel}^2$ whereas for heavy-light systems the series is an expansion in
$\alpha_s$ and $ \Lambda_{\QCD}/m$. These coefficients are the radiative corrections 
which compensate for the missing contribution of ultraviolet modes in QCD which 
are omitted by the imposition of the high-momentum ultraviolet cutoff in the lattice
formulation. QCD and the effective lattice theory agree for infrared scales sufficiently 
below the heavy quark mass $m$ and so the radiative corrections are governed by momenta larger than
$m\gtrsim 1/a$ where the strong coupling constant is small. It is therefore
legitimate to calculate these corrections in perturbation theory. The coefficients are
computed by equating the matrix elements of the operator in QCD and of its NRQCD expansion
for any choice of external states. Because the coefficients are independent of the 
matrix element chosen, we carry out the radiative matching calculation for appropriately chosen
external on-shell quark states; this is the usual technique employed in matching calculations
for NRQCD (see, for example, \cite{Morningstar:1997ep,Hart:2006ij}). These radiative corrections are 
expected to be comparable in size to higher order corrections in the $1/m$ expansion. 

Owing to the complexity of the lattice actions used, the generation of
Feynman rules has been automated
\cite{Luscher:1985wf,Hart:2004bd,Hart:2009nr} 
and the resulting integrals are calculated using the adaptive Monte Carlo integrator \vegas\
\cite{Lepage:1977sw,Lepage:1980dq}, possibly after smoothing the
integrand by adding an infrared subtraction function. In
\cite{Horgan:2009ti} radiative corrections to the heavy quark action have
been calculated for a range of frame velocities. In this work we
extend this calculation to the matching coefficients for the leading
order operators of the vector and tensor currents.

The outline of this paper is as follows. We discuss the 
different heavy-light continuum operators that contribute to 
$B\rightarrow \pi\ell\nu$ and rare $B$ decays in 
Sec.~\ref{sec:continuumoperators}, and we introduce the various quark 
actions used in this project in Sec.~\ref{sec:quarkactions}. The 
central part of this work, the matching calculation between continuum 
and lattice operators, is presented in 
Sec.~\ref{sec:matchingcalculation} where we calculate one-loop matrix 
elements both in the continuum and in the effective lattice theory and 
combine them to obtain the matching coefficients. In 
Sec.~\ref{sec:numerical_results} we present numerical results. We 
summarize and discuss our findings in Sec.~\ref{sec:summary}.

Some preliminary results from this work appeared in
Ref.~\cite{Meinel:2008th,Muller:2009af,Liu:2009dj}.

\section{Continuum operators}
\label{sec:continuumoperators}

In general we shall denote current operators by $J_n^{(\Gamma)}$ with the appropriate
number of Lorentz indices and where $\Gamma$ labels the symmetry and
transformation properties and the integer subscript, $n$, distinguishes different operators 
with the properties. 

An example is the semileptonic decay $B\rightarrow \pi\ell\nu$ which occurs 
at tree level in the quark picture. It is mediated by the hadronic vector current
\begin{equation}
  J_0^{(V)\mu} = \overline{q}_L\dgamma^\mu \Psi_L,
  \label{eqn:vectorcurrent}
\end{equation}
where $q$ and $\Psi$ are the light- and heavy-quark spinors. We follow the convention in Appendix A of Ref. \cite{Horgan:2009ti} for the Dirac gamma matrices $\dgamma^\mu$.
Only left-handed particles (denoted by the subscript $L$) participate in 
the weak interaction. For massless leptons the total width of this decay 
is proportional to the square of the form factor $f_+(q^2)$ defined by
\begin{eqnarray}
  \langle\pi(p')|J_0^{(V)\mu}|B(p)\rangle &=&
  f_+(q^2)\left(p^\mu+p'^\mu-\frac{M_B^2-M_\pi^2}{q^2}q^\mu\right)\notag\\
  &+&
  f_0(q^2)\frac{M_B^2-M_\pi^2}{q^2}q^\mu
\end{eqnarray}
and the CKM matrix element $V_{ub}$.

The situation is more complicated for rare $B$ decays which can only
occur at loop level in the Standard Model. After integrating out
physics at the electroweak scale, the transition is described by a set
of effective operators $Q_j$ with their associated Wilson coefficients
$C_j(\mu)$ \cite{Buras:1998ra}. For the radiative decay $B\rightarrow K^*\gamma$ the Hamiltonian is
\begin{equation}
  \mathcal{H} = -\frac{4G_F}{\sqrt{2}}V_{tb}V_{ts}^*\sum_j C_j(\mu) \; Q_j
  \label{eqn:effectiveconhamiltonian}
\end{equation}
where the operator basis $Q_j$ used in this work is given, for example, in
Refs.~\cite{Greub:1996tg,Ghinculov:2003qd}:
\begin{eqnarray}
  Q_1 &=& (\overline{c}_{L\beta}\dgamma^\mu b_{L\alpha})
  			(\overline{s}_{L\alpha}\dgamma_\mu c_{L\beta}), 
                         \label{eqn:introduction:effective_operators}\\
  Q_2 &=& (\overline{c}_{L\alpha}\dgamma^\mu b_{L\alpha})
  			(\overline{s}_{L\beta}\dgamma_\mu c_{L\beta}), \notag \\
  Q_3 &=& (\overline{s}_{L\alpha}\dgamma^\mu b_{L\alpha})
  			\sum_{q=u,d,c,s,b} (\overline{q}_{L\beta}\dgamma_\mu q_{L\beta}),
			\notag \\ 
  Q_4 &=& (\overline{s}_{L\alpha}\dgamma^\mu b_{L\beta})
  			\sum_{q=u,d,c,s,b} (\overline{q}_{L\beta}\dgamma_\mu q_{L\alpha}),
			\notag \\ 
  Q_5 &=& (\overline{s}_{L\alpha}\dgamma^\mu b_{L\alpha})
  			\sum_{q=u,d,c,s,b} (\overline{q}_{R\beta}\dgamma_\mu q_{R\beta}),
			\notag \\ 
  Q_6 &=& (\overline{s}_{L\alpha}\dgamma^\mu b_{L\beta})
  			\sum_{q=u,d,c,s,b} (\overline{q}_{R\beta}\dgamma_\mu q_{R\alpha}),
			\notag \\
  Q_7 &=& \frac{e}{16\pi^2} m_b \overline{s}_{L\alpha} \sigma^{\mu\nu}
				 b_{R\alpha} F_{\mu\nu},
				 \notag \\
  Q_8 &=& \frac{g}{16\pi^2} m_b \overline{s}_{L\alpha} \sigma^{\mu\nu}
				 T^a_{\alpha\beta} b_{R\beta} G_{\mu\nu}^a, \notag
\end{eqnarray}
with
\begin{eqnarray}
\sigma^{\mu\nu} = \frac{i}{2} 
				 [\dgamma^\mu,\dgamma^\nu].
\end{eqnarray}				 
$F_{\mu\nu}$ ($G^a_{\mu\nu}$) is the electromagnetic (chromodynamic) 
field strength tensor; $\alpha$ and $\beta$ are color indices.

This factorization separates the physics at large energy scales, which is 
contained in the model-dependent Wilson coefficients $C_j(\mu)$, from 
the universal hadronic matrix elements of the effective operators 
$Q_j$.

\begin{table}
  \centering
  \begin{ruledtabular}
    \begin{tabular}{cdcd}
      $C_1$ & 0.016 & $C_5$ & 0.017 \\
      $C_2$ & 0.711 & $C_6$ & 0.009 \\
      $C_3$ & -0.078 & $C_7$ & -0.300 \\
      $C_4$ & 0.093 & $C_8$ & -0.144 \\
    \end{tabular}
  \end{ruledtabular}
  \caption{Numerical values of the Standard Model Wilson coefficients
    in the effective Lagrangian (\ref{eqn:effectiveconhamiltonian})
    for the radiative decay $B\rightarrow K^*\gamma$. The coefficients
    $C_j(\mu_b)$ are evaluated in the leading logarithmic (LL)
    approximation at the scale $\mu_b=5.0\MeV$ using the so-called
    ``magic numbers'' in Ref.~\cite{Buras:1998ra} with
    $m_Z=91.1876\MeV$, $m_W=\mu_W=80.425\MeV$ and
    \mbox{$\alpha_s(m_Z)=0.118$}.
  }
  \label{tab:wilsoncoefficients}
\end{table}

\subsection{Wilson coefficients}
\label{sec:wilson_coefficients}

By matching the effective Hamiltonian in Eqn. 
(\ref{eqn:effectiveconhamiltonian}) to the Standard Model it is easy 
to see that at tree level $C_2=1$ and all other coefficients are zero. 
After summing the leading logarithms of the form $\left[\alpha_s 
\log(\mu_W/\mu_b)\right]^n$ the dominant contribution to $C_7$ comes 
from the mixing of $Q_2$ to $Q_7$ via one loop diagrams 
\cite{Buras:1998ra}. The Standard Model Wilson coefficients relevant 
for the radiative decay are given in Tab. \ref{tab:wilsoncoefficients} 
in the leading logarithmic (LL) approximation. As can be seen from 
these numbers, the dominant operators are $C_2$ and $C_7$. Numerical 
values of $C_j$ are now known at next-to-next-to-leading (NNLL) order 
in the Standard Model \cite{Misiak:2006zs}.

The Wilson coefficients are model dependent. For example, in
\cite{Grinstein:1990tj} it is shown how $C_7(\mu_W)$ and $C_8(\mu_W)$
change in a two Higgs doublet model. The numerical size of these
changes depends on the parameters of the specific model. In
\cite{Grinstein:1990tj} it is reported that the inclusive decay rate
$B\rightarrow X_s\gamma$, which in the LL approximation is
proportional to $|C_7(\mu_b)|^2$, could in principle be enhanced by about a factor of
three compared to the Standard Model.

Although this enhancement is now ruled out by recent calculations of the Standard Model branching ratio in the inclusive $B\rightarrow X_s\gamma$ \cite{Misiak:2006zs,Becher:2006pu} and exclusive $B\rightarrow K^*\gamma$ \cite{Ali:2007sj,Ball:2006eu,Becher:2005fg} decays, which  are compatible with experimental results \cite{Barberio:2008fa}, the situation is less clear for the time dependent CP asymmetry in the exclusive decay: although it is expected to be small in the Standard Model \cite{Ball:2006cva} it has not yet been measured to sufficiently high precision \cite{Aubert:2005bu,Ushiroda:2006fi}.  In the Standard Model the opposite chirality operator, which is obtained by replacing $b_R\mapsto b_L$ and $s_L\mapsto s_R$ in $Q_7$ is suppressed by a factor of $m_s/m_b$ as the spin flip requires the insertion of a mass term. This is not necessarily the case in the new physics models studied in \cite{Atwood:1997zr} where it is shown that the opposite chirality operator and mixing induced CP asymmetries can be enhanced even if the branching ratio agrees with Standard Model predictions.

Similar conclusions can be drawn for the decay \mbox{$B\rightarrow K^{(*)}\ell\ell$} \cite{Ali:1999mm,Ali:2002jg}, the forward-backward asymmetry is dependent on $C_7$, $C_9$ and $C_{10}$. The current experimental measurements \cite{Aubert:2006vb,Adachi:2008sk,Aaltonen:2008xf} will be improved by the LHCb experiment \cite{Schune:2009zz} and help constrain these coefficients.
\subsection{Local and non-local operators}
\label{sec:lattice_operators}

The operators in
Eqns.~(\ref{eqn:effectiveconhamiltonian},\ref{eqn:introduction:effective_operators}) can be split into two groups:
the four quark operators 1-6 couple two hadronic currents at one point in
space time.
All other operators couple a heavy-light quark current to a gauge
boson or a leptonic current. These two sets of operators contribute
differently to hadronic heavy-light matrix elements
\cite{Grinstein:2000pc}.

\subsubsection{Local contributions}

\begin{figure}[t]
  \centering
  \includegraphics[width=0.5\linewidth]{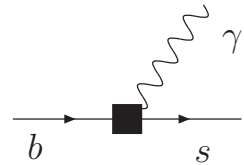}
  \caption{Local tensor operator diagram contributing to the decay
  $B\rightarrow K^*\gamma$}
  \label{fig:tensordiagram}
\end{figure}

The local contribution to the radiative decay $B\rightarrow K^*\gamma$ is described by the tensor current (see Fig.~\ref{fig:tensordiagram})
\begin{equation}
  Q_7 = \frac{e}{16\pi^2}m (\overline{q}_L \sigma^{\mu\nu} \Psi_R) F_{\mu\nu}
  \label{eqn:Q7}.
\end{equation}
The local contributions to the leptonic decay \mbox{$B\rightarrow
K^{(*)}\ell\ell$} also include operators
\begin{align}
  Q_9 &= \frac{e^2}{16\pi^2} \overline{q}_L\dgamma^\mu \Psi_L 
  \sum_\ell \overline{\ell}\dgamma_\mu \ell, 
  \nonumber\\
  Q_{10} &= \frac{e^2}{16\pi^2} \overline{q}_L\dgamma^\mu \Psi_L 
  \sum_\ell \overline{\ell}\dgamma_\mu \dgamma^5 \ell,
  \label{eqn:Q9and10}
\end{align}
which couple the heavy-light vector current $J_0^{(V)\mu}$ to a vector
(or axial vector) leptonic current.

In the following we will consider the vector current in
Eqn.~(\ref{eqn:vectorcurrent}) and the heavy-light hadronic tensor
current
\begin{equation}
  J_0^{(T)\mu\nu} =  m (\overline{q}_L\sigma^{\mu\nu} \Psi_R).
  \label{eqn:tensorcurrent}
\end{equation}
The heavy
quark mass is included in this current as only left-handed particles
participate in the weak interaction: flipping the chirality on one of
the external legs requires the insertion of a mass term
$m\overline{q}_Lq_R + (\mathrm{h.c.})$. Whilst in the Standard Model
the operator $m_q (\overline{q}_R \sigma^{\mu\nu}\Psi_L)$ with
opposite chirality is suppressed by $m_q/m$ relative to
Eqn.~(\ref{eqn:tensorcurrent}), this is not necessarily
the case in new physics models \cite{Atwood:1997zr}.

As we set the light quark mass to zero in the matching calculation we
will drop all chiral projectors in the following.

\subsubsection{Non-local contributions}

\begin{figure}[t]
  \centering
  \includegraphics[width=0.5\linewidth]{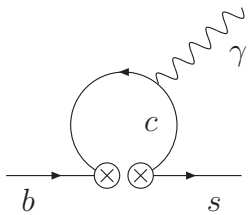}
  \caption{Non-local four quark operator diagram contributing to the
  decay $B\rightarrow K^*\gamma$}
  \label{fig:fourquarkdiagram}
\end{figure}

Non-local contributions come from diagrams like the one in 
Fig.~\ref{fig:fourquarkdiagram}: The gauge bosons couple to an internal quark loop which is created by contracting the two charm fields of a four quark 
operator. Given the size of $C_2$ it is 
important to estimate the effect of these diagrams on the hadronic 
matrix element. 

\paragraph*{Long distance effects.}
\label{sec:long_distance}

\begin{figure}
  \centering
  \includegraphics[width=0.9\linewidth]{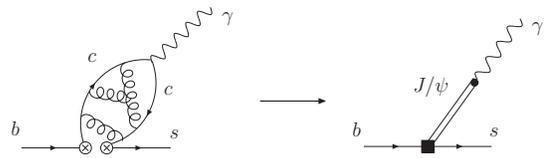}
  \caption{Resonant contribution to $b\rightarrow s\gamma$ from four
  quark operators.}
  \label{fig:Jpsi}
\end{figure}

The dominant contribution to the long distance amplitude induced by
the $b\rightarrow s\overline{c}c$ operators is usually assumed to come
from the diagram where the photon couples directly to the charm quark loop.
This is confirmed by the perturbative calculation combined with a
quark model in \cite{Asatrian:1999mt} where it is found that the main
contribution generated by $Q_2$ is given by diagrams where the photon
couples to the $\overline{c}c$ loop and the gluon connects this loop
to either the $b$ or $s$ quark. Only $Q_7$ contributes at tree level
but in \cite{Asatrian:1999mt} it is argued that the $\order(\alpha_s)$
contribution of the four quark operator is of the same order as the
one loop correction to the electromagnetic tensor operator $Q_7$.

The resonant contribution from the charm loop can be described as the 
decay $B\rightarrow V\psi_n$, where the $\psi_n$ is a bound 
$\overline{c}c$ vector state, such as $J/\psi$, which subsequently 
decays into a photon, see Fig. \ref{fig:Jpsi}.

In this approximation the long-distance amplitude can be written as
\begin{align}
  A &= Q_ce\sum_{n,\epsilon_n}
  \frac{(\epsilon^*_n)_\mu\langle0|\overline{c}\dgamma^\mu
  c|\psi_n\rangle\mathcal{A}(B\rightarrow
  V\psi_n)}{q^2-M_n^2+iM_n\Gamma_n}.
\end{align}
where $\epsilon_n$ is the polarization of the vector meson $\psi_n$
with mass $M_n$ and width $\Gamma_n$. For real photons with $q^2=0$
the sum is dominated by the lowest lying resonances. The mass of the
$J/\psi$ is $3.097\GeV$ so that long distance effects from charm loops
are expected to be suppressed by the inverse of this mass.

This argument is supported by the explicit calculation in
\cite{Khodjamirian:1997tg} where the charm quark is integrated out to
obtain an effective $b\overline{s}g\gamma$ operator suppressed by
$1/m_c^2$. The matrix element of this local operator is calculated
using QCD sum rules and found to be small, contributing around $5\%$
of the dominant amplitude from $Q_7$.

\paragraph*{Chromomagnetic tensor operator.}
\label{sec:chromomagnetic_tensor}

The contribution from the chromomagnetic tensor operator is estimated
in \cite{Carlson:1994pg}. There the decay amplitude is calculated for
both the electromagnetic and the tensor operator in the framework of a
quark model. The contribution of $Q_8$ is found to be suppressed
relative to $Q_7$ by a factor $\Lambda_{\operatorname{QCD}}/m_B\times
C_8/C_7 \approx 5\%$.

To summarize, matrix elements of four quark operators and of the 
chromomagnetic tensor operator are suppressed for small $q^2$ and it 
is likely that they contribute little to the radiative decay.
With currently available techniques these non-local contributions can
not be treated in lattice QCD and, if required, must be calculated
using different approaches, such as QCD sum rules
\cite{Cheng:1994qw,Greub:1994tb,Carlson:1994pg,Khodjamirian:1997tg,Asatrian:1999mt,Ball:2004ye,Ball:2006eu}
to complement the lattice calculation of the local operators.
In the following we will concentrate on the vector current
$\overline{q}_L\dgamma^\mu \Psi_L$ and the tensor current
$m(\overline{q}_L \sigma^{\mu\nu} \Psi_R)$ which are the hadronic parts
of the electroweak operators in (\ref{eqn:Q7}) and
(\ref{eqn:Q9and10}). Nonperturbative lattice matrix elements elements
of these currents have been calculated \cite{Meinel:2008th,Liu:2009dj}
by using mNRQCD as an effective theory for the heavy quark.
\section{Lattice quark actions}
\label{sec:quarkactions}

On currently available lattices the Compton wavelength of the $b$ 
quark is smaller than the lattice spacing. We discretize the heavy 
quark in an effective, non-relativistic theory where the high frequency 
fluctuations have been integrated out. The construction of this action 
is described in \cite{Horgan:2009ti} and in the following we summarize the 
main results.

The light quarks are discretized using highly improved relativistic 
actions \cite{Lepage:1998vj,Follana:2006rc}. In the one loop 
calculations presented in this paper vacuum polarization effects of 
the light quark do not contribute.

\subsection{Moving NRQCD}

The (tree level) NRQCD
action is obtained by decoupling the quark and antiquark degrees of
freedom in the fermionic action by a Foldy-Wouthuysen-Tani
transformation. The theory can be Lorentz-transformed to a moving
frame and higher order time derivatives are removed by a subsequent
field transformation. On the lattice, where Lorentz invariance is
broken by discretization, this will give rise to a new theory which is
known as m(oving) NRQCD. As the theory contains only first order time
derivatives, propagators can be computed very efficiently by a single
sweep through the lattice.

The mNRQCD action is given by
\begin{equation}
  S = \sum_{\vec{x},\tau}\psi^\dagger (\vec{x},\tau)\left[
  \psi(\vec{x},\tau) - K(\tau)\psi(\vec{x},\tau-1)\right]
\end{equation}
with kernel
\begin{multline}
  K(\tau) = \left( 1-\frac{\delta H}{2} \right)
  \left( 1-\frac{H_0}{2n} \right)^n
  U_4^\dagger(\vec{x},\tau-1)
  \\
  \times\left( 1-\frac{H_0}{2n} \right)^n
  \left( 1-\frac{\delta H}{2} \right).
  \label{sec:HQ_lattice:NRQCDkernel}
\end{multline}
The lowest order kinetic term $H_0$ is
\begin{equation}
  H_0 = - i \vec{v}\cdot\vec{\Delta}^\pm
  -\frac{\Delta^{(2)}-\Delta_v^{(2)}}{2\gamma m} 
  \label{eqn:HQ_lattice:mNRQCDH0}
\end{equation}
where $\vec{v}$ is the frame velocity and $\Delta_j^\pm$,
$\Delta^{(2)}$ and $\Delta_v^{(2)}$ are first and second order gauge
covariant finite difference operators defined in Ref.~\cite{Horgan:2009ti}.
In momentum space the non-relativistic dispersion relation
\begin{align}
  E &= \vec{v}\cdot\vec{p}_{\res} + 
  \frac{\vec{p}_{\res}^2-(\vec{v}\cdot\vec{p}_{\res})^2}{2\gamma m}+\dots 
\nonumber \\ 
&\approx \sqrt{m^2+(m\gamma \vec{v}+\vec{p}_{\res})^2}-\gamma m\nonumber\\
&= \sqrt{m^2+\vec{p}^2}-\gamma m
\end{align}
is obtained. $\delta H$ contains higher order corrections in $1/m$ and
operators which remove discretization errors. The action used in this
work is described in Ref.~\cite{Horgan:2009ti}; it is correct to
$\order(1/m^2,v_{\rel}^4)$, where $v_{\rel}$ is the relative velocity
of the two quarks in a heavy-heavy system.

The integer stability parameter $n$ is introduced to remove numerical
instabilities for smaller quark masses $m$.

\subsection{Relativistic quark actions}
\label{sec:relquarkactions}

We separately consider two different staggered lattice actions
describing the light quarks.

The \asqtad\ action \cite{Lepage:1998vj} suppresses ``taste-breaking''
interactions of lattice doublers by $\order(\alpha_s a^2)$. This is
done by introducing form factors for one gluon emission in the action.
The Highly Improved Staggered Quark (\hisq) action reduces
discretization errors further by an additional level of smearing
followed by link unitarization
\cite{Follana:2006rc}.

Physically, the light quark mass is much smaller than the hadronic
scale and in the matching calculation it will be set to zero. This
simplifies the calculation and leads to additional relations between
different matching coefficients due to chiral symmetry.

The MILC and UKQCD Collaborations have produced a set of lattice
configuration ensembles including \asqtad\ vacuum polarization effects
and are currently extending this to configurations with dynamical
\hisq\ fermions
\cite{Bazavov:2009jc,Bazavov:2009wm}. 
These configurations are used in the
nonperturbative calculation of heavy-light form factors
\cite{Meinel:2008th,Liu:2009dj}.

\section{Matching calculation}
\label{sec:matchingcalculation}

To match heavy-light operators, expressed in terms of effective mNRQCD heavy
quark fields, to those in the continuum theory, we must
calculate radiative corrections both in the continuum and on the
lattice. We match the theories at leading order in the $1/m$ expansion
and one loop order in $\alpha_s$. Matrix elements of $\order(1/m)$ operators are expected to be of the same size as the leading radiative corrections and are matched at tree level.

\subsection{Continuum calculation}

In the continuum we calculate the one loop matrix elements of the
currents $J_0^{(\Gamma)}$ and expand the result in inverse powers of
the heavy quark mass. We use the notation $\Gamma = V,T$ to denote the
expressions in
Eqns.~(\ref{eqn:vectorcurrent},\ref{eqn:tensorcurrent}). Owing to
Lorentz invariance these expansions can be expressed as linear
combinations of a small number of tree level matrix elements:
\begin{equation}
  \langle q |J_0^{(\Gamma)}| b \rangle_{\mathrm{con}} =
  \sum_j Z_j^{(\Gamma,\mathrm{con})} \langle q |J_j^{(\Gamma)}|b \rangle_\mathrm{tree}.
\end{equation}
At leading order in the $1/m$ expansion the operators that contribute
to the vector current are:
\begin{align}
  J_0^{(V)\mu} &= \overline{q} \dgamma^\mu \Psi,
  \notag \\
  J_1^{(V)\mu} &= \overline{q} u^\mu \Psi
  \label{eqn_vec_cur}
\end{align}
and to the tensor current:
\begin{align}
  J_0^{(T)\mu\nu} &= m(\overline{q} \sigma^{\mu\nu} \Psi),
  \notag \\
  J_1^{(T)\mu\nu} &= 2im \left(\overline{q} 
  (\dgamma^\mu u^\nu - u^\mu \dgamma^\nu)\right)\Psi
  \label{eqn_tens_cur}
\end{align}
where $u^\mu = (\gamma,\gamma\vec{v})$ is the frame velocity. The one
loop contributions to the mixing matrix \mbox{$Z_j^{(\Gamma,\mathrm{con})} =
\delta_{j0} + \alpha_s \delta Z_j^{(\Gamma,\mathrm{con})} + \dots$} are
for the vector operator \cite{Morningstar:1997ep}
\begin{align}
  \delta Z_0^{(V,\mathrm{con})} &= \frac{1}{3\pi} \left(
  -\frac{11}{4}- \frac{3}{2}\log \lambda^2/m^2 \right),
  \nonumber \\
  \delta Z_1^{(V,\mathrm{con})} &= \frac{2}{3\pi},
  \label{eqn:mixingmatrixvector}
\end{align}
and for the tensor operator
\begin{align}
  \delta Z_0^{(T,\mathrm{con})} &= \frac{1}{3\pi} \left(
    -\frac{27}{4}-\frac{3}{2}\log\lambda^2/m^2 + 4\log m^2/\mu^2
  \right),
  \nonumber \\
  \delta Z_1^{(T,\mathrm{con})} &= 0.
  \label{eqn:mixingmatrixtensor}
\end{align}
We introduced a gluon mass $\lambda$ to regulate infrared divergences.
It is important to use the same infrared regulator in both continuum
QCD and the effective lattice theory, any dependence on the gluon mass
will cancel in the matching coefficients.

As the tensor current is not conserved, its anomalous dimension does
not vanish:
\begin{equation}
  \gamma_T^{(\mathrm{con})} = \frac{8\alpha_s}{3\pi} + \dots.
\end{equation}
The coefficients of the infrared logarithms in Eqns. 
(\ref{eqn:mixingmatrixvector}) and (\ref{eqn:mixingmatrixtensor}) 
agree due to heavy quark symmetry.

\subsection{Construction of lattice operators}

On the lattice we must construct operators $J_0^{(\Gamma,\mathrm{lat})}$
which have the same on-shell matrix elements as the associated continuum
operators:
\begin{equation}
  \langle q|J_0^{(\Gamma,\mathrm{lat})}|b\rangle_{\mathrm{lat}} =
  \langle q|J_0^{(\Gamma)}| b\rangle_{\mathrm{con}}.
\end{equation}
At tree level, the operators in the effective theory are obtained from
Eqns.~(\ref{eqn:vectorcurrent},\ref{eqn:tensorcurrent}) by applying
the field transformation
\begin{equation}
  \Psi(x) = S(\Lambda)\tilde{T}(\tilde{x}) e^{-im\dgamma^0 u\cdot x}
  A_{D_t} \frac{1}{\sqrt{\gamma}}\Psi_v.
\end{equation}
$S(\Lambda)$ is a spinorial Lorentz boost, $\tilde{T}$ the FWT
transformation decoupling the quark- and antiquark fields in the rest
frame and $A_{D_t}$ an additional field transformation to remove
higher order time derivatives. \mbox{$\Psi_v(x) =
\left(\psi_v(x),0\right)^T$} is the (positive energy) field in the
effective theory.

Using the explicit expressions in Ref.~\cite{Horgan:2009ti} one finds at
$\order(1/m)$
\begin{equation}
  \Psi = \frac{1}{\sqrt{\gamma}} \left(
  1 - \frac{i\dgamma_0 \vec{v}\cdot\vec{D}}{2m} + 
  \frac{i\vec{\dgamma}\cdot\vec{D}}{2m}
  + \frac{i\vec{v}\cdot\vec{D}}{2\gamma m}
  \right)S(\Lambda) \Psi_v
\end{equation}
from which the tree level currents in the effective theory can be read
off by inserting the field transformation in
Eqns.~(\ref{eqn:vectorcurrent},\ref{eqn:tensorcurrent}).

In the following we calculate the one loop matching coefficients of
the leading order operators in the $1/m$ expansion. At this order both
$\tilde{T}$ and $A_{D_t}$ are equal to the identity. The Lorentz boost
is
\begin{equation}
   S(\Lambda) = \frac{1}{\sqrt{2(1+\gamma)}} \left(
    (1+\gamma) - \gamma v \hat{\vec{v}}\cdot\vec{\dgamma}\dgamma^0
  \right).
\end{equation}
We choose $\hat{\vec{v}}$, the direction of the frame velocity, to be 
along one of the lattice axes, i.e. $\hat{\vec{v}} = (1,0,0)$. We can 
then classify the lattice directions (and associated Lorentz indices) 
as timelike (denoted 0, as usual), parallel to $\hat{\vec{v}}$ 
(denoted $\parallel$) or perpendicular to $\hat{\vec{v}}$ (denoted 
$\perp$).

\subsubsection{Choice of operator basis}

In the continuum the operator basis $J_0^{(\Gamma)}$, $J_1^{(\Gamma)}$
is used. On the lattice Lorentz invariance is broken and it is
convenient to work in another basis which is spanned by operators with
different Dirac structure. Firstly, the operators $J_{0}^{(\Gamma)}$ and $J_{1}^{(\Gamma)}$ are
split into the sum of two operators
$J_n^{(\Gamma)}=J_{n,1}^{(\Gamma)}+J_{n,2}^{(\Gamma)}$. For $n=0$:
\begin{align}
  J_{0,1}^{(\Gamma)} &= \rho^{(\Gamma)}f_1(v)\overline{q}(x) 
  \Gamma \Psi_v,
  \nonumber \\
  J_{0,2}^{(\Gamma)} &= -\rho^{(\Gamma)}f_2(v)\overline{q}(x) 
  \Gamma \vec{\hat{v}}\cdot\vec{\dgamma}\dgamma^0\Psi_v \; .
  \label{eqn:bsgamma_matching:lattice_operators}
\end{align}
In the vector case $\Gamma=\dgamma^\mu$ and $\rho^{(V)} = 1$, and
in the tensor case $\Gamma=\sigma^{\mu\nu}$ and $\rho^{(T)} = m$. The
velocity dependence has been absorbed in the functions
\begin{align}
  f_1(v) &= \sqrt{\frac{1+\gamma}{2\gamma}}, &
  f_2(v) &= v\sqrt{\frac{\gamma}{2(1+\gamma)}} \; .
  \label{eqn_velocity_fns}
\end{align}
The corresponding leading order (in the heavy quark expansion)
operators $J_{1,j}^{(\Gamma)}$ are obtained by replacing \mbox{$\Gamma \mapsto
u^\mu$} in the vector case, and $\Gamma \mapsto 2i(\dgamma^\mu u^\nu -
\dgamma^\nu u^\mu)$ for the tensor.

The tree level matrix elements of all these operators form a basis for
expanding higher order matrix elements. In the following we do not, however, need to
include $J_{1,j}^{(\Gamma)}$, because the tree level matrix elements
are given by linear combinations of those of $J_{0,j}^{(\Gamma)}$ due to heavy quark symmetry $\dgamma^0 \Psi_v=\Psi_v$. In
the vector case:
\begin{align}
  J_{1,1}^{(V)0} &= \gamma J_{0,1}^{(V)0}, &
  J_{1,2}^{(V)0} &= -\gamma J_{0,2}^{(V)0},
  \nonumber \\
  J_{1,1}^{(V)\parallel} &= (1+\gamma) J_{0,2}^{(V)\parallel}, &
  J_{1,2}^{(V)\parallel} &= (1-\gamma) J_{0,1}^{(V)\parallel},
  \label{eqn:bsgamma_matching:operatordecomposition:vector}
\end{align}
and for the tensor operator:
\begin{align}
  J_{1,1}^{(T)0\parallel} &= 2\left(
    \gamma J_{0,1}^{(T)0\parallel}
    +(1+\gamma) J_{0,2}^{(T)0\parallel}
  \right), \nonumber\\
  J_{1,2}^{(T)0\parallel} &= 2\left(
    (1-\gamma)J_{0,1}^{(T)0\parallel}
    - \gamma J_{0,2}^{(T)0\parallel}
  \right),
  \nonumber \\
  J_{1,1}^{(T)0\perp} &= 2\gamma J_{0,1}^{(T)0\perp}, \nonumber\\
  J_{1,2}^{(T)0\perp} &= -2\gamma J_{0,2}^{(T)0\perp},
  \nonumber \\
  J_{1,1}^{(T)\parallel\perp} &= -2(1+\gamma) J_{0,2}^{(T)\parallel\perp}, 
  \nonumber\\
  J_{1,2}^{(T)\parallel\perp} &= -2(1-\gamma) J_{0,1}^{(T)\parallel\perp}.
  \label{eqn:bsgamma_matching:operatordecomposition:tensor}
\end{align}
Clearly, this decomposition is not Lorentz invariant, but can be
carried out for fixed frame velocity.

On the lattice the two operators in
Eqn.~(\ref{eqn:bsgamma_matching:lattice_operators}) mix under
renormalization,
\begin{multline}
  \langle q|J_{0,j}^{(\Gamma)}|b\rangle_{\lat} =
  \sum_k \left(\delta_{jk}+\alpha_s\delta 
  Z^{(\Gamma,\lat)}_{jk} + \dots \right) \times
  \\
  \langle q |J_{0,k}^{(\Gamma)}| b\rangle_{\tree} \; ,
  \label{eqn:bsgamma_matching:one_loop_lattice}
\end{multline}
(with $j,k = 1,2$). Instead of using this basis (which we will
call the $(1,2)$~basis), it is more convenient to work in the
$(+,-)$~basis:
\begin{equation}
J_{0,\pm}^{(\Gamma)} = J_{0,1}^{(\Gamma)}\pm J_{0,2}^{(\Gamma)}, 
\end{equation}
as only $J_{0,+}^{(\Gamma)}$ contributes to processes at tree level. We may
then write
\begin{multline}
  \langle q|J_{0,+}^{(\Gamma)}|b\rangle_{\con} =
  (1+\alpha_s \delta Z_+^{(\Gamma,\con)}) 
  \langle q |J_{0,+}^{(\Gamma)}| b\rangle_{\tree} + 
  \\
  \alpha_s \delta Z_-^{(\Gamma,\con)} 
  \langle q |J_{0,-}^{(\Gamma)}| b\rangle_{\tree}.
  \label{eqn:bsgamma_matching:one_loop_continuum}
\end{multline}
On the lattice renormalization factors $\delta Z_{jk}^{(\Gamma,\lat)}$ in this
basis can then be defined in an analogous way to
Eqn.~(\ref{eqn:bsgamma_matching:one_loop_lattice}) [i.e. where
$j,k=\pm$].
For the vector operator, we must distinguish whether the Lorentz index
of the current is timelike, parallel or perpendicular to the frame velocity.
Relations like Eqn.~(\ref{eqn:bsgamma_matching:operatordecomposition:vector})
can then be used to relate $Z_\pm^{(V,\con)}$ to $Z_0^{(V,\con)}$ and $Z_1^{(V,\con)}$:
\begin{align}
  \delta Z_+^{(V,\con)0} &= \delta Z_0^{(V,\con)},
  \notag\\
  \delta Z_-^{(V,\con)0} &= \gamma \delta Z_1^{(V,\con)},
  \notag \\
  \delta Z_+^{(V,\con)\parallel} &= 
  \delta Z_0^{(V,\con)}+\delta Z_1^{(V,\con)},
  \notag \\
  \delta Z_-^{(V,\con)\parallel} 
  &= -\gamma\delta Z_1^{(V,\con)},
  \notag\\
  \delta Z_+^{(V,\con)\perp} 
  &= \delta Z_0^{(V,\con)}
  \notag,\\
  \delta Z_-^{(V,\con)\perp} &= 0 \;.
\end{align}

For the tensor operator there is no dependence on the Lorentz indices:
\begin{align}
  \delta Z_{+}^{(T,\con)} &= \delta Z_0^{(T,\con)},
  \nonumber \\
  \delta Z_{-}^{(T,\con)} &= 0 \; .
\end{align}
\subsubsection{Matching coefficients}

Combining
Eqns.~(\ref{eqn:bsgamma_matching:one_loop_lattice},\ref{eqn:bsgamma_matching:one_loop_continuum}),
the lattice operator which has the same one loop matrix elements as
the continuum operator is
\begin{equation}
  J_0^{(\Gamma,\lat)} = \left(1+\alpha_s 
  c_+^{(\Gamma)} \right)J_{0,+}^{(\Gamma)} + 
  \alpha_s c_-^{(\Gamma)} J_{0,-}^{(\Gamma)}
\end{equation}
with
\begin{align}
  c_+^{(\Gamma)} &= \delta Z_+^{(\Gamma,\con)} - 
  \delta Z_{++}^{(\Gamma,\lat)},
  \nonumber \\
  c_-^{(\Gamma)} &= \delta Z_-^{(\Gamma,\con)} - 
  \delta Z_{+-}^{(\Gamma,\lat)} \; .
\end{align}
\subsection{Mixing matrix}

In the $(1,2)$~basis of operators, the lattice mixing matrix can be split into
a diagonal part and a contribution $\xi_{jk}^{(\Gamma)}$ from one
particle irreducible (1PI) diagrams,
\begin{equation}
  \delta Z_{jk}^{(\Gamma,\lat)} = \left(
    \delta Z_{\mathrm{mult}}^{(\Gamma)}
     - \delta Z_{f_j}
  \right) \delta_{jk}
  +\xi_{jk}^{(\Gamma)}.
\end{equation}
For the vector current the multiplicative renormalization contains the
wavefunction renormalization only: \mbox{$\delta Z_{\mathrm{mult}}^{(V)} =
\frac{1}{2}\left(\delta Z_q + \delta Z_\psi\right)$}. For the tensor
current, however, there is an additional contribution from the
renormalization of the heavy quark mass: \mbox{$\delta
Z_{\mathrm{mult}}^{(T)} = \frac{1}{2}\left(\delta Z_q + \delta
Z_\psi\right)-\delta Z_m$}.

The relation between renormalized and bare parameters is $\vec{v}_R =
Z_v v$, $\gamma_R = (1-\vec{v}_R^2)^{-1/2}$, $m_R = Z_m m$ and $q =
\sqrt{Z_q} q_R$, $\Psi = \sqrt{Z_\psi} \Psi_R$. $m_R$ is the pole mass, which can be defined perturbatively both in the continuum and on the lattice. The renormalization
constants can be expanded perturbatively using the generic formula $Z_x
= 1 +\alpha_s \delta Z_x+\dots$.

The renormalization of the velocity functions $f_{1,2}$ in
Eqn.~(\ref{eqn_velocity_fns}) is $f_{j,R} = Z_{f_j} f_j$ with
\begin{align}
  \delta Z_{f_1} &= \frac{1-\gamma}{2}\delta Z_v, & 
  \delta Z_{f_2} &= \frac{1+\gamma}{2}\delta Z_v.
\end{align}
One then finds 
\begin{align}
  \delta Z_{++}^{(\Gamma,\lat)} &= \delta Z_{\mathrm{mult}}^{(\Gamma)} 
 -\frac{1}{2}\delta Z_v + \xi_{++}^{(\Gamma)},
  \nonumber \\
  \delta Z_{+-}^{(\Gamma,\lat)} &= \frac{\gamma}{2}\delta Z_v + 
  \xi_{+-}^{(\Gamma)}.
\end{align}
Even though we use only the leading order heavy-light operators we
still include $1/m$ corrections in the action. Next we isolate
infrared divergences in the renormalization constants and find (in
Feynman gauge)
\begin{align}
  \delta Z_q &= \frac{1}{3\pi}\log a^2\lambda^2 + F_{q},
  \nonumber \\
  \delta Z_\psi &= -\frac{2}{3\pi}\log a^2\lambda^2 + F_{\psi}(v,am),
  \notag\\
  \delta Z_m &= F_m(v,am),
  \notag\\
  \delta Z_v &= F_v(v,am),
  \notag\\
  \xi_{++}^{(\Gamma)} &= -\frac{1}{3\pi}
  \log a^2\lambda^2+F^{(\Gamma)}_{\xi_{++}}(v,am)
  ,\notag\\
  \xi_{+-}^{(\Gamma)} &= F^{(\Gamma)}_{\xi_{+-}}(v,am).
  \label{eqn:bsgamma_matching:IRdivergences}
\end{align}
Here we make the lattice spacing $a$ explicit and $\lambda$ is the
gluon mass used in the continuum calculation. The infrared divergence
of $\xi_{++}^{(\Gamma)}$ is independent of the Dirac structure
$\Gamma$ due to heavy quark symmetry and can be inferred from the
subtraction integral discussed in
Sec.~\ref{sec:IRsubtractionfunction}. The functions $F_x$ are infrared
finite and can, if required, be expanded in powers of the inverse
heavy quark mass on the lattice.

It is interesting to note that the logarithms in
Eqn.~(\ref{eqn:bsgamma_matching:IRdivergences}) represent both
infrared ($\lambda\rightarrow0$) and ultraviolet ($a\rightarrow0$)
divergences. A similar combination of short and long distance
divergences occurs in HQET if the theory is regulated in dimensional
regularization: The heavy quark propagator does not contain any scales
and the integral vanishes in this case (see, for example,
Ref.~\cite{Eichten:1990vp}). In the HQET case,
this simplifies the calculation of the matching coefficients as only
the QCD integrals need to be calculated.

\begin{widetext}

\subsection{Results for the vector operator}

After cancelling infrared divergences the final expression for the
matching coefficients $c_\pm^{(\Gamma)}$ is
\begin{align}
  c_+^{(V)0} &= -\frac{11}{12\pi} - \frac{1}{2}\left(F_q + F_\psi\right)
  + \frac{1}{2}F_v
  + \frac{1}{2\pi}\log a^2m^2-F_{\xi_{++}}^{(V)0}, 
  \notag\\
  c_+^{(V)\parallel} &= -\frac{1}{4\pi} - 
  \frac{1}{2}\left(F_q + F_\psi\right)\notag+\frac{1}{2}F_v
  +\frac{1}{2\pi}\log a^2m^2-F_{\xi_{++}}^{(V)\parallel}, 
  \notag\\
  c_+^{(V)\perp} &= -\frac{11}{12\pi} - 
  \frac{1}{2}\left(F_q + F_\psi\right)+\frac{1}{2}F_v
  + \frac{1}{2\pi}\log a^2m^2-F_{\xi_{++}}^{(V)\perp}, 
  \notag\\
  c_-^{(V)0} &= \frac{2\gamma}{3\pi}-
  \frac{\gamma}{2}F_v-F_{\xi_{+-}}^{(V)0},
  \notag\\
  c_-^{(V)\parallel} &= -\frac{2\gamma}{3\pi}- 
  \frac{\gamma}{2}F_v-F_{\xi_{+-}}^{(V)\parallel}, 
  \notag\\
  c_-^{(V)\perp} &= -\frac{\gamma}{2}F_v-F_{\xi_{+-}}^{(V)\perp}.
  \label{eqn:bsgamma_matching:c++full_vector}
\end{align}
In the limit $v\rightarrow0$ the operator $J_{0,2}^{(\Gamma)}$ does not contribute as it is proportional to $v$; in NRQCD there is only one operator with matching coefficient $c^{(\Gamma)}$:
\begin{equation}
  c_1^{(\Gamma)} = c_+^{(\Gamma)} + c_-^{(\Gamma)} \rightarrow c^{(\Gamma)}
  \qquad\text{for}\quad v\rightarrow 0.
\end{equation}
We find
\begin{align}
  c^{(V)0} &= -\frac{1}{4\pi} -\frac{1}{2}\left(F_q + F_\psi\right) + 
  \frac{1}{2\pi}\log a^2m^2-F_{\xi}^{(V)0}, 
  \notag\\
  c^{(V)j} &= -\frac{11}{12\pi} -  \frac{1}{2}\left(F_q + F_\psi\right) + 
  \frac{1}{2\pi}\log a^2m^2-F_{\xi}^{(V)j}.
\end{align}

The matching coefficient of the zero component of the vector (or axial
vector) current at $v=0$ has been calculated in \cite{Dalgic:2003uf} and the corresponding calculation for the spatial components can be found in \cite{Dalgic:2006dt}. We tested our code by reproducing these results.

\subsection{Results for the tensor operator}

The corresponding results for the tensor operator are:
\begin{align}
  c_+^{(T)\mu\nu} &=
  -\frac{9}{4\pi}-\frac{1}{2}\left(F_q+F_\psi\right)
  +F_m
  +\frac{1}{2}F_v - F^{(T)\mu\nu}_{\xi_{++}}
  +\frac{1}{2\pi}\log a^2m^2 + \frac{4}{3\pi}\log m^2/\mu^2
  \notag,\\
  c_-^{(T)\mu\nu} &= -\frac{\gamma}{2}F_v -F^{(T)\mu\nu}_{\xi_{+-}}
  \label{eqn:bsgamma_matching:c++full_tensor}
\end{align}
and in the NRQCD limit $v=0$
\begin{equation}
  c^{(T)\mu\nu} =
  -\frac{9}{4\pi}-\frac{1}{2}\left(F_q+F_\psi\right)
    +F_m-F^{(T)\mu\nu}_{\xi} +
  \frac{1}{2\pi}\log a^2m^2 + \frac{4}{3\pi}\log m^2/\mu^2.
\end{equation}

\end{widetext}
\subsection{The anomalous dimension}

The ultraviolet behavior of the lattice theory is described by the
logarithmic terms in
Eqns.~(\ref{eqn:bsgamma_matching:c++full_vector},\ref{eqn:bsgamma_matching:c++full_tensor}).
In particular, the $\log a^2m^2$ term is a UV divergence which is
independent of the Dirac structure of the renormalized operator due to
heavy quark symmetry. As the short distance behavior of the effective
theory is different from that of continuum QCD, its coefficient is not
the same as that of the $\log m^2/\mu^2$ term in
Eqn.~(\ref{eqn:bsgamma_matching:c++full_tensor}). The anomalous
dimension of the lattice operator can be obtained by noting that the
renormalized operator is related to the bare operator by
multiplication by $Z_{\Gamma}^{(\lat)}$:
\begin{equation}
  J_{0,+}^{(\Gamma,\mathrm{ren})} = \left(Z_\Gamma^{(\lat)}\right)^{-1} 
  J_{0,+}^{(\Gamma)}.
\end{equation}
The counterterm has to be chosen such that it absorbs the logarithmic
UV divergence in $\delta Z_{++}^{(\Gamma,\lat)}$,
\begin{equation}
  Z_\Gamma^{(\lat)} = 1 - \frac{\alpha_s}{2\pi}\left[\log a^2\mu^2_{\lat} + 
    (\text{finite terms})\right] + \dots
\end{equation}
where $\mu_{\lat}$ is an arbitrary scale which cancels in
physical results. We thus find
\begin{equation}
  \gamma_\Gamma^{(\lat)} = \frac{1}{Z_\Gamma^{(\lat)}}\;\frac{dZ_\Gamma^{(\lat)}}
       {d\log \mu_{\lat}} = -\frac{\alpha_s}{\pi} + \dots.
\end{equation}
This agrees with the result for HQET regularized in dimensional
regularization \cite{Manohar:2000dt}.

\subsection{Quark renormalization parameters}
\label{sec:lightquarkrenparm}

The wavefunction renormalization of massless \asqtad\ quarks has been
calculated to one loop in \cite{Dalgic:2003uf}. We repeated this calculation
with a larger number of points in the \vegas\ integration to obtain
\begin{equation}
  \delta Z_q^{\operatorname{ASQTad}} = -0.92411(42) + 
  \frac{1}{3\pi}\log a^2\lambda^2
\end{equation}
in Feynman gauge (the error is the statistical error of the \vegas\ integral). For massless \hisq\ quarks we obtain
\begin{equation}
  \delta Z_q^{\operatorname{HISQ}} = -0.3905(16) + 
  \frac{1}{3\pi}\log a^2\lambda^2.
\end{equation}
The one loop renormalization of the heavy quark wavefunction, frame 
velocity and mass is given for a range of frame velocities in 
Ref.~\cite{Horgan:2009ti}. As demonstrated there the magnitude of all renormalization parameters 
is reduced significantly by including mean fields corrections. Only 
the wavefunction renormalization has a logarithmic IR divergence, 
given by $-\frac{2}{3\pi}\log a^2\lambda^2$.

\subsection{One particle irreducible matrix elements}
\label{sec:onePI}

\begin{figure}[t]
  \centering
  \includegraphics[width=0.6\linewidth]{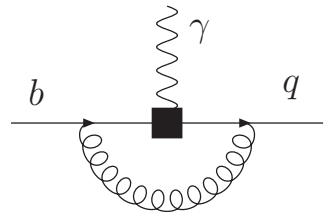}
  \caption[One particle irreducible diagram]{One particle irreducible
  diagram at one loop}
  \label{fig:bsgamma_matching:1PIdiagram}
\end{figure}

The one particle irreducible (1PI) matrix elements can be found by
evaluating the one loop diagram in
Fig.~\ref{fig:bsgamma_matching:1PIdiagram}.

The $\order(\alpha_s)$ one particle irreducible correction to
the operators $J_{0,j}^{(\Gamma)}=f_j \overline{q}\Gamma S_j
\Psi_v$ (with \mbox{$S_1=1$}, \mbox{$S_2=-\vec{\dgamma}\cdot\hat{\vec{v}}\dgamma_0$}) is given by
\begin{align}
  \langle q |J_{0,j}^{(\Gamma)}| b\rangle_{\mathrm{lat,1PI}} &= 
  \alpha_s f_j \overline{u}(p') \Sigma_j^{(\Gamma)} U
  \notag \\
  &= \alpha_s \sum_{k=1,2} 
  \xi_{jk}^{(\Gamma)} \langle q |J_{0,k}^{(\Gamma)}|b\rangle_{\mathrm{tree}} 
  \notag\\
  &= \alpha_s
  \sum_{k=1,2} \xi_{jk}^{(\Gamma)} f_k \overline{u}(p') \Gamma S_k U
  \label{lattice:eqn:1PIdecomposition}
\end{align}
where the heavy-quark four spinor is
\begin{equation}
  U^{(\sigma)} = \begin{pmatrix} \chi^{(\sigma)} \\ 0 \end{pmatrix}, 
  \quad \text{with} \quad
  \chi^{(\sigma)} \in \left\{ 
  \begin{pmatrix}1\\0\end{pmatrix},
    \begin{pmatrix}0\\1\end{pmatrix}
      \right\}.
\end{equation}
$\Sigma_j^{(\Gamma)}$ is the lattice integrand, after factoring out external quark spinors, the strong coupling constant and the velocity function $f_j$.
To extract $\xi_{jk}^{(\Gamma)}$, we replace the spinors by Euclidean
\footnote{In the rest of this section we work with Euclidean vectors and gamma matrices, which are related to their counterparts in Minkowski space by \mbox{$x_0=x^0 = ix_{(M)}=ix_0^{(0)}$},
\mbox{$x_j=x^j = x_{(M)}^j=-x_j^{(M)}$},
\mbox{$\gamma_0=\gamma^0 = \dgamma^0=\dgamma_0$},
\mbox{$\gamma_j=\gamma^j = -i\dgamma^j = i\dgamma_j$}.
}
on-shell projection operators:
\begin{align}
  \overline{u}(p')\mapsto \Pi_q(p') & \equiv
  \sum_{\sigma=\spinup,\spindown} u^{(\sigma)}(p') \overline{u}^{(\sigma)}(p')
  = -i\pslash',
  \nonumber \\
  U\mapsto\Pi_b & \equiv \sum_{\sigma=\spinup,\spindown} U^{(\sigma)} U^{(\sigma)T}
  = \frac{1}{2}(1+\gamma_0)
\end{align}
and take the trace of (\ref{lattice:eqn:1PIdecomposition}):
%
 \begin{align}
  & f_j
  \mathrm{Tr}\left[
     \Pi_q(p') \Sigma_j^{(\Gamma)} \Pi_b \Pi^{(\Gamma)}
  \right] \notag\\
  &= \sum_{k=1,2} \xi_{jk}^{(\Gamma)} f_k
  \mathrm{Tr}\left[
    \Pi_q(p') \Gamma S_k \Pi_b \Pi^{(\Gamma)}
  \right] \label{lattice:eqn:1PIdecompositionTraced}
 \end{align}
%
The Dirac matrix $\Pi^{(\Gamma)}$ is a suitable projection operator
which depends on $\Gamma$; as both sides of (\ref{lattice:eqn:1PIdecompositionTraced}) 
are a (linear) function of the four momentum $p'$, this relation defines $\xi_{jk}^{(\Gamma)}$ 
for all $j,k$.

\subsubsection{Infrared subtraction function}
\label{sec:IRsubtractionfunction}
In some configurations in momentum space the integration contour in 
the $k_0$ plane is pinched by poles. This leads to large peaks in the 
integrand and can generate infrared divergences in the final result.

We construct an appropriate infrared subtraction function
$f^{(\mathrm{sub})}$ to smooth the integrand and thus speed up the
convergence of the \vegas\ estimate of the integral. The 1PI matrix
elements can be written as
\begin{eqnarray}
  \xi_{jk}^{(\Gamma)} &=& \int \dd{k} \left(f^{(\Gamma,\mathrm{lat})}_{jk}-f^{(\mathrm{sub})}_{jk}\right) 
  + \int \dd{k} f^{(\mathrm{sub})}_{jk} \notag\\
  &=& \xi^{(\Gamma,\mathrm{lat})}_{jk} + \xi^{(\mathrm{sub})}_{jk}.
\end{eqnarray}
Construction of the subtraction function is guided by the continuum
integral, which has the same infrared behavior as the corresponding
lattice expression. In the continuum the 1PI correction to the operator
$J_{0,j}^{(\Gamma)}$ at the matching point $p'=0$, $p=(m,0)$ is given by
\begin{eqnarray}
  & \int \dd{k} \left[\overline{u}(p') (-igT^a \dgamma_\rho)
    \frac{-i\kslash}{k^2} f_j
    \Gamma S_j U\right] \times
  \notag \\
  &\qquad D^{(0)}_{h}(k)(-gT^av_\rho) \frac{1}{k^2+\lambda^2}
  \notag \\
  &\equiv \alpha_s f_j \; \overline{u}(p')\Sigma_j^{(\Gamma,\mathrm{sub})} U
  \label{subtraction:eqn:eucl_subtraction_integral}
\end{eqnarray}
where $D_h^{(0)}(k)$ is the heavy quark propagator at \mbox{$p=(m,0)$}. This integral can be rendered UV finite 
without changing the infrared structure by replacing
\begin{equation}
  D^{(0)}_{h}(k) = \frac{-i}{k_0-i\vec{v}\cdot\vec{k}} \mapsto
  \frac{2\gamma m}{(k+m u)^2+m^2},
\end{equation}
where the metric is $g_{\mu\nu} = \operatorname{diag}(+1,+1,+1,+1)$ in Euclidean space, and
the velocity four-vector is given by $u=(i\gamma,\gamma\vec{v})$. As in Eqn.~(\ref{lattice:eqn:1PIdecomposition}) 
we write
\begin{widetext}
  \begin{align}
    f_j \overline{u}(p') \Sigma_j^{(\Gamma,\mathrm{sub})}U & = 
    \frac{16\pi}{3} f_j
    \int\dd{k} \overline{u}(p')(-i\dgamma_\rho) \frac{-i\kslash}
    {k^2}\Gamma S_j U
    \frac{2\gamma m}{(k+p)^2+m^2}
    (- v_\rho) \frac{1}{k^2+\lambda^2}
    \nonumber \\
    & = \sum_{k=1,2} f_k \xi^{(\mathrm{sub})}_{jk}\overline{u}(p')\Gamma S_k U.
  \end{align}
\end{widetext}
The subtraction integral is independent of the Dirac structure and is
easy to solve analytically:
\begin{align}
  \xi_{jk}^{(\mathrm{sub})} &= \int \dd{k} f^{(\mathrm{sub})}_{jk} 
  \nonumber \\ 
  &= -\frac{\delta_{jk}}{3\pi} 
  \left( 1 + \log \lambda^2/m^2 \right) + \order(\lambda/m).
\end{align}
This concludes our discussion of the structure of the matching calculation. In the next section we present numerical values for different quark masses and frame velocities.
\section{Numerical results}\label{sec:numerical_results}

\begin{figure}[t]
  \centering
  \includegraphics[height=0.9\linewidth,angle=270]{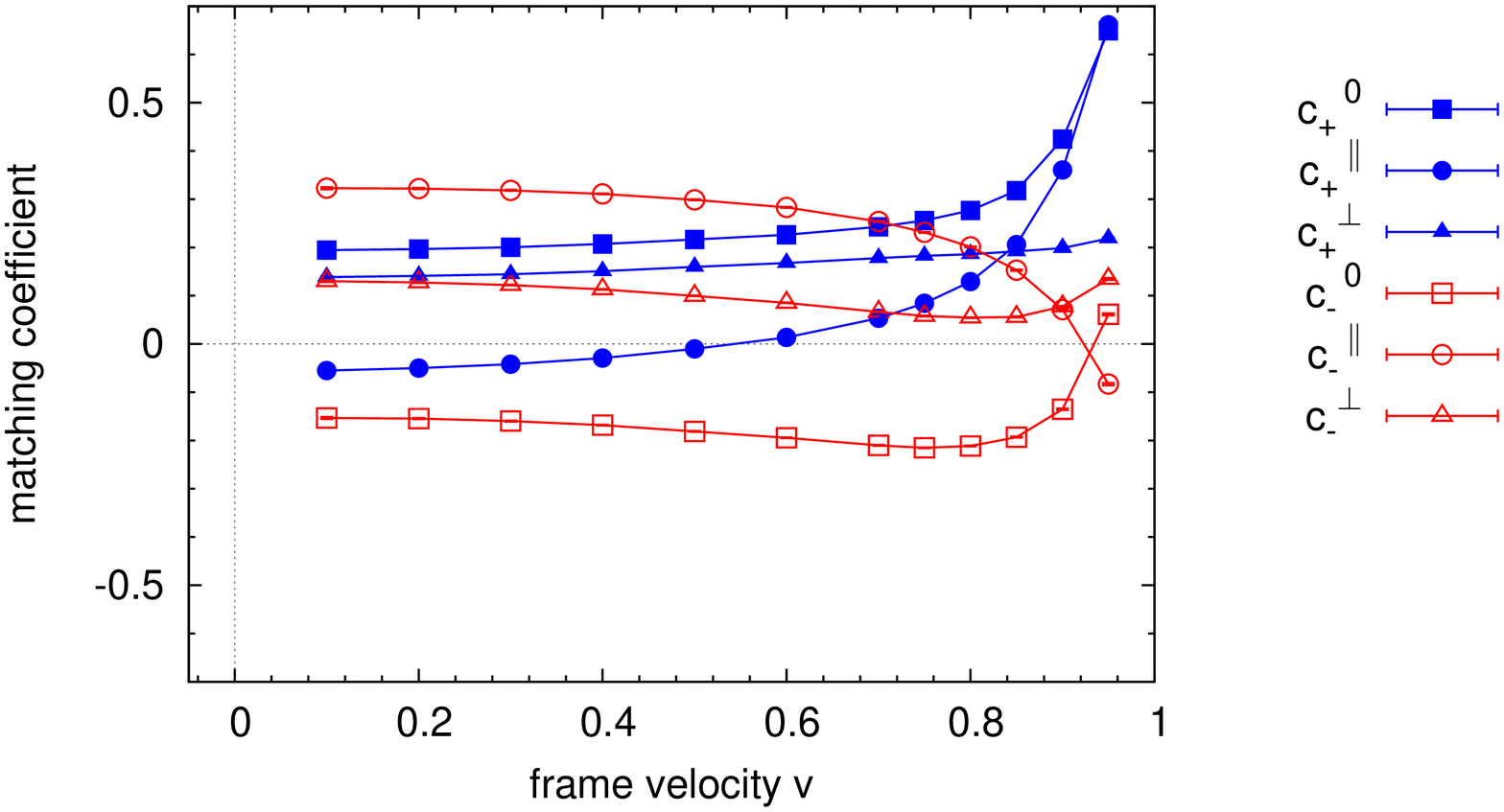}
  \\[4ex]
  \includegraphics[height=0.9\linewidth,angle=270]{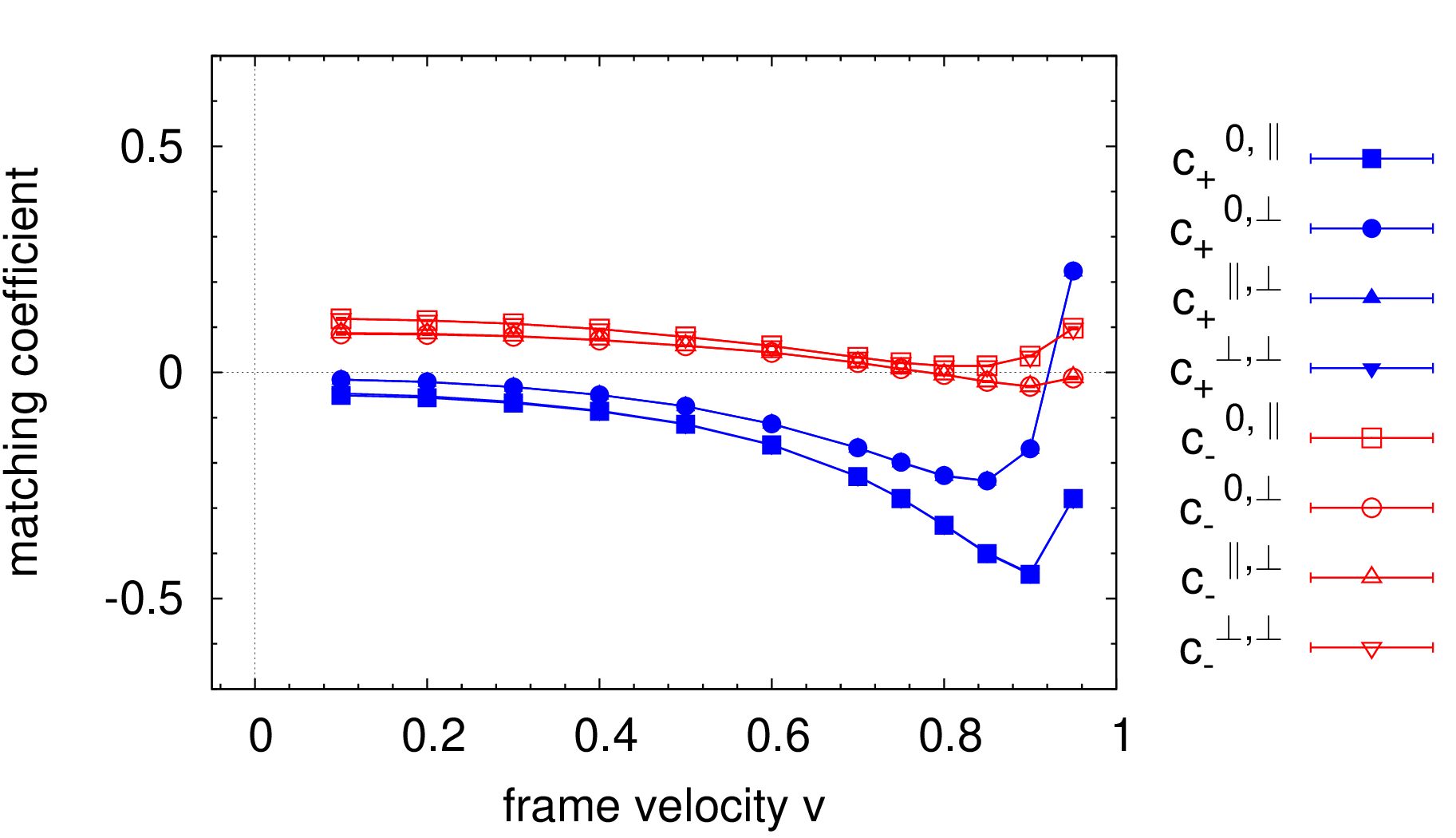}
  \caption{Matching coefficients, vector current (top) and tensor
  current (bottom). The heavy quark mass is \mbox{$m=2.8$} and the \asqtad\ action is used to discretize the
  light quark. Due to chiral symmetry discussed in the main text the following matching coefficients are identical for the tensor current:
  $c_-^{0\parallel}$ (open red squares)
  and
  $c_-^{\perp\perp}$ (open red downward triangles), 
  $c_-^{0\perp}$ (open red circles)
  and
  $c_-^{\parallel\perp}$ (open red upward triangles), 
  $c_+^{0\parallel}$ (filled blue squares)
  and
  $c_+^{\perp\perp}$ (filled blue downward triangles), 
  $c_+^{0\perp}$ (filled blue circles)
  and
  $c_+^{\parallel\perp}$ (filled blue upward triangles).}
  \label{fig:bsgamma_matching:c_match_full}
\end{figure}

\begin{figure}[t]
  \centering
  \includegraphics[height=0.9\linewidth,angle=270]{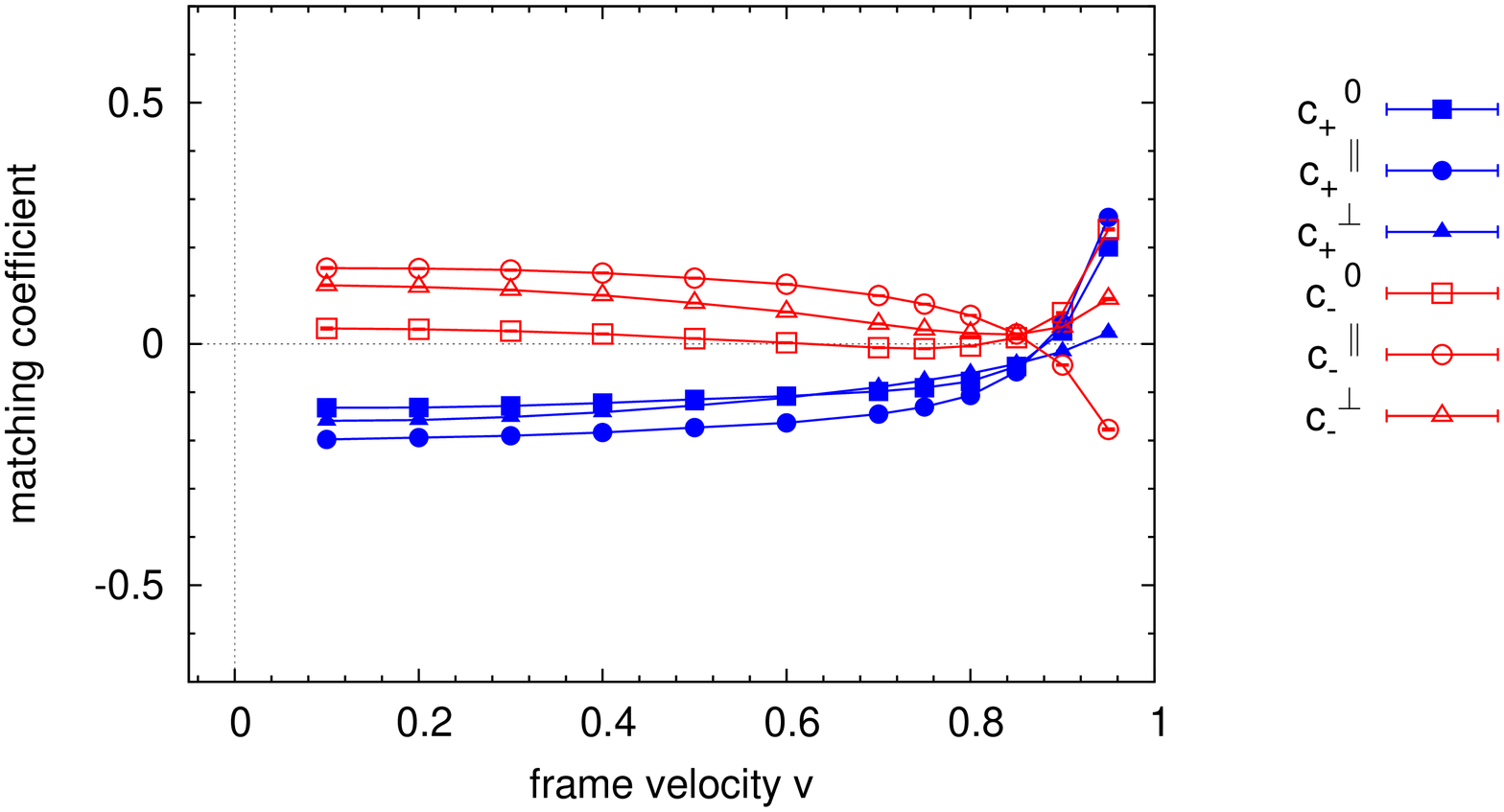}
  \\[4ex]
  \includegraphics[height=0.9\linewidth,angle=270]{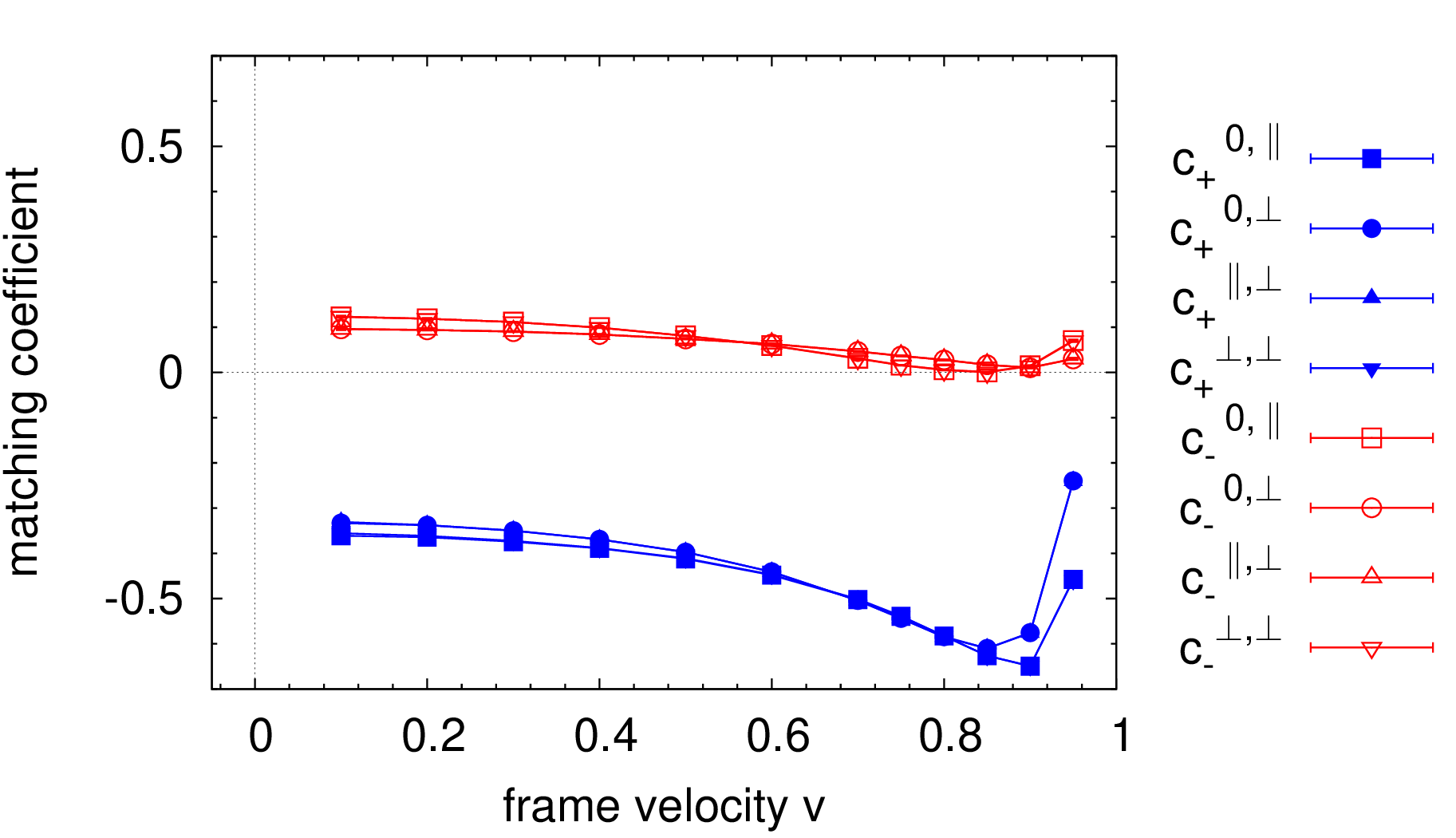}
    \caption{Matching coefficients, vector current (top) and tensor
    current (bottom). The heavy quark mass is \mbox{$m=2.8$} and the \hisq\ 
    action is used to discretize the
    light quark. See the comment on Fig. \ref{fig:bsgamma_matching:c_match_full} for relations between the tensor current matching coefficients.}
  \label{fig:bsgamma_matching:c_match_full_hisq}
\end{figure}

In Figs.~\ref{fig:bsgamma_matching:c_match_full}
and~\ref{fig:bsgamma_matching:c_match_full_hisq} we show results for
the matching coefficients for the vector and tensor current (see
Tables~\ref{tab:bsgamma_matching:c_match_full_vector}
to~\ref{tab:bsgamma_matching:c_match_full_HISQ} for numerical values).

In both cases we use a heavy mass of $m=2.8$ and a stability parameter
$n=2$. These are the values currently used in nonperturbative
calculations of heavy-light form factors on coarse MILC lattices
\cite{Meinel:2008th,Liu:2009dj}. The gluon action is Symanzik improved
\cite{Lepage:1994yd}. We present results both for the \asqtad\ and
\hisq\ light quark action.

For the vector current we calculate the matching coefficients for
three different directions of the Lorentz index $\mu$: temporal
($\mu=0$), parallel to the frame velocity (denoted $\parallel$) and
perpendicular ($\perp$). The frame velocity is chosen to be along the
lattice axis $\mu=1$. For $v=0$ we consider a corresponding set of
directions: $\mu=0$, $\mu=1$ and $\mu=2$.

For the tensor current there are four different cases for indices
$(\mu,\nu)$: $(0,\parallel)$, $(0,\perp)$, $(\parallel,\perp)$ and
$(\perp,\perp)$. For $v=0$ we choose $(\mu,\nu)=(0,1)$, $(0,2)$,
$(1,2)$ and $(2,3)$.
The renormalization scale of the tensor current is $\mu=m$.

\subsection{Discussion}

As the light quark is massless its propagator and vertex functions anticommute with $\dgamma^5$. With $\sigma^{01} = i\dgamma^5 
\sigma^{23}$, this implies that matching coefficients for $(0,\parallel)$ and 
$(\perp,\perp)$ are identical (given that we boost in the direction 
$\mu=1$). The same holds for $(0,\perp)$ and 
$(\parallel,\perp)$, as $\sigma^{02} = -i\dgamma^5 \sigma^{13}$. For $v=0$ the matching coefficients for all combinations of $(\mu,\nu)$ 
agree as there is no preferred direction. Heavy quark symmetry ($\dgamma^0\tilde{\Psi}_v=\tilde{\Psi}_v$) and relations such as $\dgamma^0\dgamma^j = -i \sigma^{0j}$ can be used to relate 1PI matrix elements of the vector and tensor current.

We emphasize that the magnitude of all matching coefficients is 
reduced by including mean field corrections in the renormalization 
parameters, the dependence on the frame velocity is weak and all 
matching coefficients (except for $c_+^{(T)}$, which depends on the renormalization scale) are of order $0.3$ or smaller for moderate $v$.
With $\alpha_s(2/a) \approx 0.3$ the relative size of the radiative 
corrections does not exceed $10\%$. From heavy quark power counting we expect the matrix elements of the $O(1/m)$ operators (which are matched at tree level) to be suppressed by the same factor $\Lambda_{\QCD}/m\approx 0.1$ relative to the leading operators.

The matching coefficients appear to diverge for \mbox{$v\rightarrow1$}. It should be noted that whereas mNRQCD reduces to NRQCD in the limit $v\rightarrow 0$, the collinear theory at \mbox{$v=1$} is qualitatively different, in particular both mass and velocity renormalization are not defined in this limit.

In addition, for large $v$ the momentum distribution of the heavy quark in the initial state meson is boosted by a factor of $\gamma$ so that for highly relativistic frame velocities the power counting in $1/m$ will break down. As shown in \cite{Horgan:2009ti} the statistical errors of nonperturbative matrix elements grow with decreasing $q^2$ and in practise it is unlikely that they will be calculated for $v\gtrsim 0.5$.
\subsubsection{Vector current}

The matching coefficient for the zero component of the vector current at $m=2.8$, \mbox{$c^{(V)0}=0.04293(52)$} in
Table~\ref{tab:bsgamma_matching:c_match_full_vector}, is in perfect
agreement with the corresponding value \mbox{$\tilde{\rho}_0=0.043(2)$} in
Table~III of \cite{Dalgic:2003uf}. For $v=0$ we find that the matching
coefficients \mbox{$c^{(V)1}=0.26970(40)$} and \mbox{$c^{(V)2}=0.26929(39)$} agree within errors as would be expected from rotational invariance and are consistent with $\tilde{\rho}_k^{(0)}=0.270(1)$ in Table~II of \cite{Dalgic:2006dt}. 

The splitting between the matching coefficients for different Lorentz
indices is reduced by using the \hisq\ light quark action. This
reduction is not, however, as pronounced as for the tensor current. In the continuum the matching coefficient $c_-^{(V)}$ is zero. Using the HISQ action reduces the magnitude of $c_-^{(V)\parallel}$ by nearly a factor of two and an even stronger reduction is observed for $c_-^{(V)0}$.

\subsubsection{Tensor current}

For the tensor current we find that the splitting between the matching
coefficients for $(\mu,\nu) = (0,\parallel)$ and $(0,\perp)$ is
reduced by using the \hisq\ action for the light quark. A similar reduction is seen in
the splitting between the $(\parallel,\perp)$ and $(\perp,\perp)$
matching coefficients.

The matching coefficients
$c_-^{(T)\mu\nu}$ are always very small and their magnitude is about $0.1$. The size of $c_+^{(T)\mu\nu}$ depends on the continuum renormalization
scale $\mu$; for $\mu=m$, we find that these coefficients are also
very small when using \asqtad\ light quarks. They are, however, larger
when the \hisq\ action is used to discretize the light quark.

\subsubsection{Quark mass dependence}
We repeated the calculations for a heavy quark mass $m=1.9$, which corresponds roughly to the bare quark mass on the fine MILC lattices. The results for a range of frame velocities are shown in Tables ~\ref{tab:bsgamma_matching:c_match_full_vector_m=1.9} to~\ref{tab:bsgamma_matching:c_match_full_HISQ_m=1.9}. The absolute size of the matching coefficients is larger than for $m=2.8$ but still typically lies in the range 0.1 - 0.5 for the frame velocities considered.

\section{Conclusion}
\label{sec:summary}

M(oving) NRQCD is a useful tool for extending the range of accessible
$q^2$ in a lattice calculation of heavy-light form factors. The
formalism has been improved and tested extensively over the last
years. Radiative corrections to the effective actions have been
calculated in a previous publication \cite{Horgan:2009ti}. Further
reduction of systematic uncertainties is justified by an increase in
precision of experimental results. In this paper we show how
systematic errors due to radiative corrections can be reduced by
renormalizing the heavy-light vector and tensor currents. After
cancelling infrared divergences the one loop corrections to matching
coefficients are of the order one and smaller. The results will be
used in the current calculation of nonperturbative form factors
\cite{Liu:2009dj}.

As the lattice imposes a cutoff $\sim 1/a$, operators which are formally 
suppressed by $1/m$ can ``mix down'' to the leading order operators 
\cite{Collins:2000ix}. These power law terms can be suppressed by constructing perturbatively subtracted $\order(1/m)$ operators $J^{(\Gamma)\mathrm{sub}}_k = J^{(\Gamma)}_k - \alpha_s \xi_{k0}^{(\Gamma)} J_0^{(\Gamma)}$ which do not mix down to the leading operators at one loop order. $\xi_{k0}^{(\Gamma)}$ is calculated from the one particle irreducible corrections to the $\order(1/m)$ operators. We find that at $v=0.4$ the non-perturbative matrix element of the subtracted $1/m$ operators is a factor of around 0.05 smaller than the leading order matrix element, which is consistent with heavy quark power counting where $\Lambda_{\QCD}/m \sim 0.1$. Before subtraction the ratio of the matrix elements can be as large as 0.3.

In this work we used the ASQTad and HISQ actions to discretize the 
light quark but it should be noted that our approach is easily 
extended to other discretizations such as the Domain Wall fermion 
action \cite{Kaplan:1992bt}.

With currently available techniques only matrix elements of local
operators in rare exclusive decays can be computed in lattice QCD.
Often these operators describe the dominant effects. For the radiative
decay $B\rightarrow K^*\gamma$ the four quark operators
$Q_3,\dots,Q_6$ are suppressed by their small Wilson coefficients. At
the physical point where $q^2=0$, the contribution of $Q_2$ is
suppressed as the $c\overline{c}$ vector resonance which connects this
operator and the external photon is far off shell. Model calculations
show that the matrix elements of the chromomagnetic operator $Q_8$ are
small. This implies that the dominant contribution comes from the
electromagnetic tensor operator $Q_7$ and the matrix elements of this
operator can be evaluated in lattice QCD.

An additional complication is that the effective heavy quark theory is
only valid at maximum recoil, i.e. at \mbox{$q^2=m^2$}. Results have to be
extrapolated to $q^2=0$ using a phenomenological ansatz. A simple,
physically motivated parametrization is given in
\cite{Becirevic:1999kt,Becirevic:2006nm}. Recently a model independent
parametrization, which uses the analyticity of the form factors, has
been suggested in \cite{Bailey:2008wp}.

Even if the dominant contribution to a given process is not given by a
local operator, lattice calculations are still useful when combined
with other approaches such as QCD sum rules.
\section*{Acknowledgements}
\label{sec:acknowledgements}

We would like to thank Christine Davies, Georg von Hippel, Lew
Khomskii, Zhaofeng Liu, Stefan Meinel, Junko Shigemitsu and Matthew
Wingate for useful discussions.

This work has made use of the resources provided by
the Darwin Supercomputer of the University of Cambridge High
Performance Computing Service (\url{http://www.hpc.cam.ac.uk}),
provided by Dell Inc.\ using Strategic Research Infrastructure Funding
from the Higher Education Funding Council for England, and the Edinburgh
Compute and Data Facility (\url{http://www.ecdf.ed.ac.uk}), which is
partially supported by the eDIKT initiative
(\url{http://www.edikt.org.uk}); and the Fermilab Lattice Gauge Theory
Computational Facility (\url{http://www.usqcd.org/fnal}).
We thank the DEISA Consortium (\url{http://www.deisa.eu}), co-funded
through the EU FP6 project RI-031513 and the FP7 project RI-222919,
for support within the DEISA Extreme Computing Initiative.
AH thanks the U.K.\ Royal Society for financial support. This work was
supported in part by the Sciences and Technology Facilities Council, SPG grant number 
PP/E006957/1. The University of Edinburgh is supported in part by the Scottish
Universities Physics Alliance (SUPA).

\appendix

\section{Pole shift}

In this appendix we discuss the choice of contour for the lattice
integrals.
As for the self-energy calculations in Ref.~\cite{Horgan:2009ti}, care has
to be taken when choosing the integration contour in the $k_0$ plane
(where $k$ is the momentum of the gluon in the loop). The heavy quark
pole must lie inside the integration contour (in the $z=e^{ik_0}$
plane) but for certain values of the loop momentum it lies outside the
unit circle. The integration contour then has to be deformed to ensure
that the result can be Wick-rotated back to Minkowski space.

In the following we discuss the corresponding contour shift for the lattice three point functions. 
We begin by mapping the positions of the poles of the light quark
propagator and then discuss the contour choice for the full
integrands.

\subsection{Poles of the light quark action}
\label{app:improved_asqtad_poles}

We use the \asqtad\ and \hisq\ actions to describe the light,
relativistic quarks and, as discussed in the main text, we treat these
quarks as being massless. The propagators for these actions are
identical, with denominator
\begin{align}
  \Delta &= 
  \sum_\nu \sin^2(k_\nu) \left(1+\frac{1}{6}\sin^2(k_\nu)\right)^2
  \notag \\
  &= \omega(1+\frac{1}{6}\omega)^2 + s(\vec{k}),
\end{align}
where \mbox{$s(\vec{k}) \equiv \sum_{j=1}^3
\sin^2(k_j)\left(1+\frac{1}{6}\sin^2(k_j)\right)^2$} and \mbox{$\omega=\sin^2(k_0)$}.

The poles of the propagator correspond to \mbox{$\Delta=0$}. For given fixed,
spatial three-momentum $\vec{k}$ (with \mbox{$k_j \in [-\pi,\pi]$}), finding
the poles reduces to solving a cubic equation with real coefficients,
\begin{equation}
  y^3 - 12y+36 s(\vec{k}) - 16 = 0,
\end{equation} 
where $y = \omega+4$. This equation either has three real
solutions, or one real solution and one conjugate pair of complex
solutions, depending on the sign of the discriminant $D$
\cite{bronstein:1996}:
defined by
\begin{align}
  D &= q^2 - 64\qquad\text{with}\quad q = 18s(\vec{k})-8 \; .
\end{align}
It has three real solutions if $D \le 0$ (or equivalently
\mbox{$s(\vec{k}) \le 8/9$}):
\begin{align}
  y_1 &= -2 P \cos \beta, &
  y_{2,3} &= 2P \cos\left(\beta \pm \frac{\pi}{3}\right)
\end{align}
where $\beta = \frac{1}{3}\arccos (q/P^3)$ and $P = 2
\operatorname{sgn} q = \pm 2$.

Alternatively, it has one real solution and one conjugate pair of
complex solutions if $D > 0$ (or equivalently $s(\vec{k}) > 8/9$):
\begin{align}
  y_1 &= - 2 P \cosh \beta, &
  y_{2,3} &= P (\cosh \beta \pm i\sqrt{3}\sinh \beta)
\end{align}
where $\beta = \frac{1}{3} \operatorname{arccosh} (q/P^3)$ and
$P$ is defined as above.

In either case, for each $\omega = y-4$, there are four
solutions for $z=e^{ik_0}$, which can be labelled as:
\begin{equation}
  z_{\pm\pm} = \pm \sqrt{1-2\omega \pm 2\sqrt{\omega^2-\omega}}.
\end{equation}
Of these twelve poles, two are ``physical'' and survive in the
continuum limit. They can be identified as those that lie on the unit
circle $|z|=1$ for $\vec{k} \to 0$. The other ten ``spurious'' poles are
lattice artifacts with masses proportional to $a^{-1}$ and therefore
decouple in the continuum limit. These poles (sometimes
called ghost poles) are, of course, not the lattice doublers.

\begin{figure}[t]
  \centering
  \includegraphics[width=0.9\linewidth]{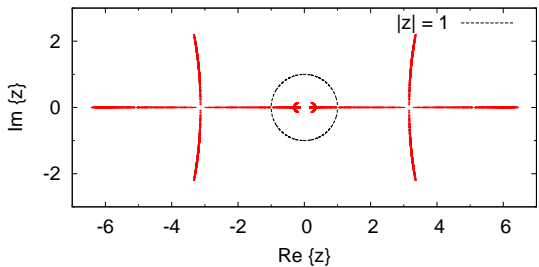}
  \caption{Poles of the massless \asq/\hisq\ fermion propagator in the
    complex plane.}
  \label{fig:poles:asqtad_poles}
\end{figure}

We calculated $z$ for a large number of random momenta $\vec{k}$ with
$k_j\in[-\pi,\pi]$; the resulting distribution of the poles in the
complex plane is shown in Fig.~\ref{fig:poles:asqtad_poles}.

\begin{figure}[t]
  \centering
  \includegraphics[width=0.9\linewidth]{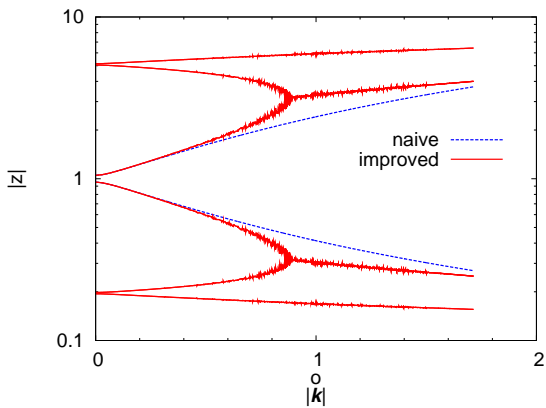}
  \caption{Absolute value of poles in the massless naive and
    \asq/\hisq\ (improved) fermion propagator as a function of
    $|\ocirc{\vec{k}}| = \sqrt{\sum_{j=1}^3 \sin^2(k_j)}$. The plot
    is generated by finding the poles for a large random sample of
    momenta $k_j \in [-\pi,\pi]$.}
  \label{fig:poles:asqtad_abs_poles}
\end{figure}

It is useful to compare the positions of these poles of the improved
actions with those of the naive propagator. The equation describing
the position of the poles is linear in $\omega$, leading to four
solutions for $z$:
\begin{equation}
  z^{(\mathrm{naive})}_{\pm\pm} = \pm\sqrt{
    1+2\ocirc{\vec{k}}^2 \pm 2\sqrt{\ocirc{\vec{k}}^2(1+\ocirc{\vec{k}}^2)}}
  \label{poles:eqn:asqtad_naive_poles}
\end{equation}
with $\ocirc{\vec{k}}^2 = \sum_{j=1}^3 \sin^2(k_j)$. We show the
comparison as a function of $|\ocirc{\vec{k}}|$ in
Fig.~\ref{fig:poles:asqtad_abs_poles}. For each spatial momentum,
\textit {all} the poles $z$ in the improved propagator lie
\textit{outside} the region defined by $|z^{(\mathrm{naive})}_{\pm+}|
< |z| < |z^{(\mathrm{naive})}_{\pm-}|$. This allows us to use the
positions of the naive poles as a safe lower bound for the positions
of the improved poles when choosing integration contours.

To take care of causality one can employ the $i\varepsilon$
prescription, changing the denominator to
\begin{equation}
  \Delta = \sum_\nu \sin^2(k_\nu) \left(1+\frac{1}{6}\sin^2(k_\nu)\right)^2
  -i\varepsilon.
\end{equation}
For the discussion of the effects of this, it is sufficient to
concentrate on the three poles outside the unit circle with positive
real part (the other poles are related to these by transformations $z
\rightarrow 1/z, -z, -1/z$). One of these is the physical pole, which
for $s(\vec{k}) = 0$ lies on the unit circle with a small negative
imaginary part. In addition, there are two additional, spurious poles
with larger real part: one with a negative (and small) imaginary part
and one with a positive (and small) imaginary part.

\begin{figure}[t]
  \centering
  \includegraphics[width=0.9\linewidth]{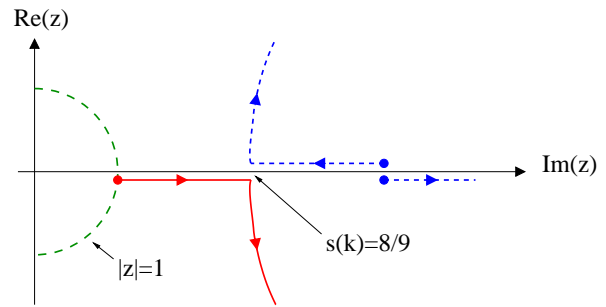}
    \caption{Movement of three of the twelve poles in the massless
      \asq/\hisq\ propagator in the $z$ complex plane (the other poles
      are related by the transformations $z \rightarrow 1/z, -z,
      -1/z$). The arrows show the motion as $s(\vec{k})$ increases.
      The physical pole is shown with a solid (red) line. The spurious
      poles are shown using a dashed (blue) line. The unit circle is
      also shown (in green). The contour pinch occurs for
      $s(\vec{k})=8/9$.}
    \label{fig:poles:asqtad_basic_poles}
\end{figure}

The movement of these poles as the momentum increases is shown in
Fig.~\ref{fig:poles:asqtad_basic_poles}.  As $s(\vec{k})$ gets larger,
the physical pole moves outwards. The spurious poles, meanwhile, move
in opposite directions: one moves outwards just below the real axis
and away from the physical pole; the other, meanwhile, moves inwards
just above the real axis, towards the physical pole.

When $s(\vec{k}) = 8/9$, the physical pole touches (within
distance $2\varepsilon$) one of the spurious poles and both, now being
complex conjugates of each other, start to move away from the real
axis in opposite directions.

Having established the positions of the poles, in the next section we
will use these to choose the contours in the lattice integration appropriately.

\subsection{Pole shift in one particle irreducible integrals}

In this section we discuss the poles of one particle irreducible three
point integrals in Sec.~\ref{sec:onePI}.

Let the position of the heavy quark pole in the $z$ plane be denoted
by $z_h$ and the poles of the naive gluon propagator by $z_\pm$ such that
$|z_-| < 1 < |z_+|$. The two poles of the naive light quark action are
$z_\pm^{(\ell)}$, whereas the six poles of the improved light quark
action are located at $z_{\pm,j}^{(\ell)}$ (and the corresponding
positions with opposite sign), ordered such that
\begin{equation}
  |z_{-,3}^{(\ell)}| < 
  |z_{-,2}^{(\ell)}| \le
  |z_{-,1}^{(\ell)}| < 1 < 
  |z_{+,1}^{(\ell)}| \le 
  |z_{+,2}^{(\ell)}| <
  |z_{+,3}^{(\ell)}| \; .
\end{equation}
Note that, as discussed in the previous section, only one of the poles is physical.

From the calculation of heavy quark renormalization parameters, it is
known for naive (Wilson) glue that $|z_h| < z_+$
\cite{Horgan:2009ti}. 
The poles of the Symanzik-improved gluon action lie outside the band
defined by $z_- < |z| < z_+$, so the same holds for improved gluons.
We can therefore concentrate on the relative positions of the poles of
the heavy and (improved) light quark propagator.

\begin{figure}[t]
  \centering
  \includegraphics[height=0.9\linewidth,angle=270]{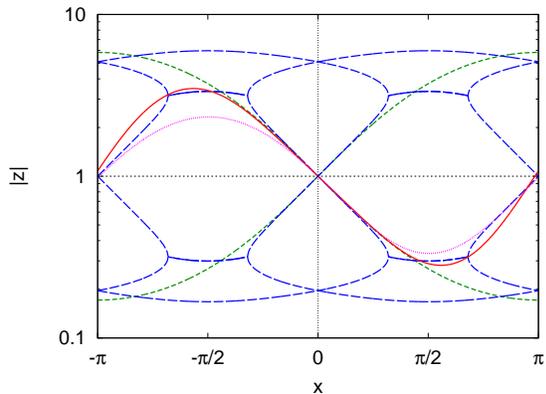}
  \caption[Absolute values of poles as a function of the loop
  momentum]{Absolute values of poles as a function of $x$ with
  $\vec{k} = (x,0,0)$. Poles of naive gluons are shown with a dashed
  (green) line; of improved fermions with a long dashed (blue) line; of the
  simple mNRQCD action ($H_0$ only) with a dotted (magenta)
  line; and of the full mNRQCD action with a solid (red)
  line. The frame velocity is $v=0.95$, the heavy quark mass $m=2.8$
  and the stability parameter $n=2$.}
\label{fig:bsgamma_matching:polecrossing}
\end{figure}

At high frame velocities and for certain choices of spatial momentum,
it turns out that the heavy quark pole can cross poles of the light
propagator outside the unit circle. Note, however, that as discussed in the main text, it is unlikely that very large frame velocities will be used in the evaluation of non-perturbative matrix elements. Examples are shown in
Fig.~\ref{fig:bsgamma_matching:polecrossing}, where we choose $\vec{k}
= (x,0,0)$ with $-\pi < x \ < \pi$ and discuss both a simple action with $H_0$ only and the full mNRQCD action.

\begin{figure*}[t]
  \centering
  \includegraphics[width=0.9\linewidth]{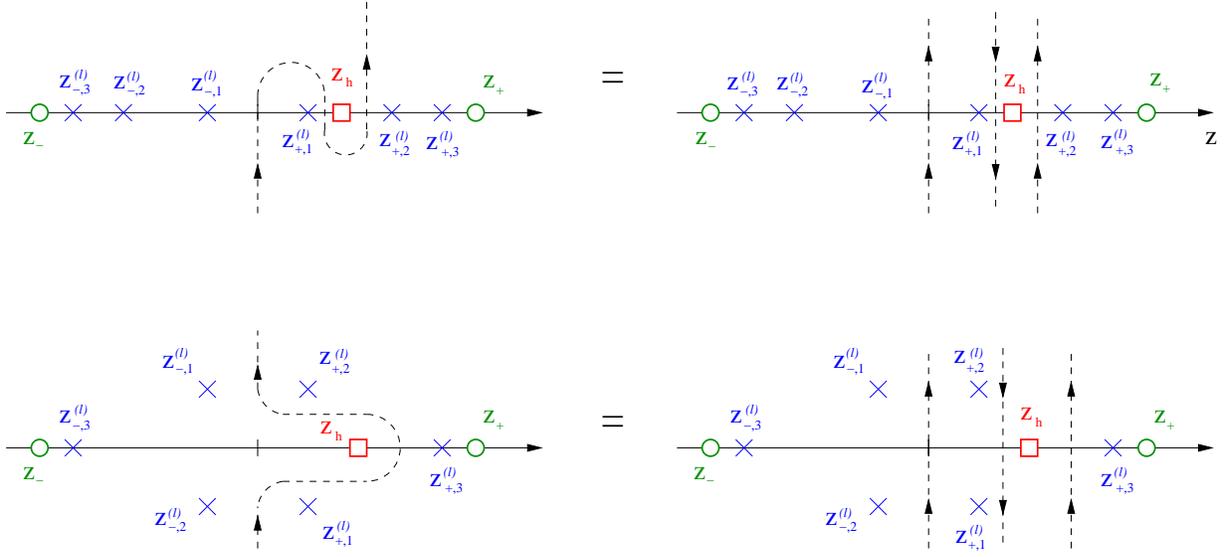}
  \caption[Integration contour for configurations with pole
      crossing]{Integration contour for momentum space configurations
      with pole crossing. The light quark poles are denoted by
      \blue{$\times$}, the naive gluon poles by \green{$\circ$} and the
      heavy quark pole by
      \red{$\square$}.}\label{fig:bsgamma_matching:triple_contour}
\end{figure*}

The crossings are seen for certain negative values of $x$, where
$|z_{+,1}^{(\ell)}| < |z_h| < |z_{+,2}^{(\ell)}|$. The problem gets
worse if the full mNRQCD action is used.
To be able to Wick-rotate back to Minkowski space in these cases, the contour needs to be deformed such that it encloses the heavy quark pole
but not the light quark poles outside the unit circle. Suitable
contours are shown on the left in
Fig.~\ref{fig:bsgamma_matching:triple_contour}. In a similar way to
that used in Ref.~\cite{Hart:2006ij}, the contours can be deformed
to avoid the poles as much as possible, arising at the triple contours
shown on the right in Fig.~\ref{fig:bsgamma_matching:triple_contour}.

Computationally, the procedure is as follows. We choose the contour(s)
separately for each value of the spatial momentum (generated, for
instance, by the \vegas\ integration code):

\begin{enumerate}
  \item $|z_h| < z_-,|z_{-,1}^{(\ell)}|$: As $z_h$ is not the smallest
  negative pole and the contour does not need to be shifted from $|z|=1$.

  \item $z_-,|z_{-,1}^{(\ell)}| < |z_h| < z_+,|z_{+,1}^{(\ell)}|$. The contour
    is shifted outwards to halfway between $|z_h|$ and
    $\min\{z_+,z_{+,1}^{(\ell)}\}$.

  \item $|z_{+,j}^{(\ell)}| < |z_h| < |z_{+,j+1}^{(\ell)}|,z_+$ for
    $j=1,2$ (see Fig.~\ref{fig:bsgamma_matching:polecrossing}). A pole
    crossing has occurred and it is necessary to integrate along three
    contours: (a) anticlockwise without shift, $|z|=1$; (b) clockwise,
    shifting the contour midway between $|z_{+,j}^{(\ell)}|$ and $|z_h|$; and
    (c) counterclockwise with the contour between $|z_h|$ and
    $\min\{|z_{+,j+1}^{(\ell)}|,z_+\}$.
\end{enumerate}

To speed up the \vegas\ calculation, in cases~(1) and~(2) we first check
using the poles of the naive light quark action, only calculating the
poles of the improved version if that test is inconclusive. When
shifting a contour midway between poles $z_a$ and $z_b$, we shift the
contour to $\sqrt{z_a z_b}$.

As the pole crossing only occurs for large momenta this is a lattice
artifact which would disappear in the continuum limit. It must,
however, be included in a lattice--continuum matching calculation.


\begin{table*}
\renewcommand{\arraystretch}{1.3}
\begin{center}
\providecommand{\pos}{{\color{white}{-}}}
\begin{tabular}{p{1cm}p{2.5cm}p{2.5cm}p{2.5cm}}
\hline\hline&&&\\[-3ex]
$v$ &\hspace{0.8cm}$c^{(V)0}$ &\hspace{0.8cm}$c^{(V)1}$ &\hspace{0.8cm}$c^{(V)2}$\\&&&\\[-3ex]\hline
$0.00$ & $\pos0.04293(52)$ &$\pos0.26970(40)$ &$\pos0.26929(39)$\\
\hline\hline
\end{tabular}
\\[4ex]
\begin{tabular}{p{1cm}p{2.5cm}p{2.5cm}p{2.5cm}}
\hline\hline&&&\\[-3ex]
$v$ & \hspace{0.8cm}$c_+^{(V)0}$ &\hspace{0.8cm}$c_+^{(V)\parallel}$ &\hspace{0.8cm}$c_+^{(V)\perp}$\\&&&\\[-3ex]\hline
$0.10$ & $\pos0.1945(21)$ &$-0.0553(22)$ &$\pos0.1384(18)$ \\
$0.20$ & $\pos0.1965(12)$ &$-0.0500(13)$ &$\pos0.1411(12)$ \\
$0.30$ & $\pos0.2005(10)$ &$-0.0419(11)$ &$\pos0.1444(10)$ \\
$0.40$ & $\pos0.20708(84)$ &$-0.0293(10)$ &$\pos0.15071(88)$ \\
$0.50$ & $\pos0.21660(77)$ &$-0.01038(88)$ &$\pos0.15972(84)$ \\
$0.60$ & $\pos0.22620(73)$ &$\pos0.01358(85)$ &$\pos0.16775(85)$ \\
$0.70$ & $\pos0.24297(71)$ &$\pos0.05349(80)$ &$\pos0.17817(87)$ \\
$0.75$ & $\pos0.25596(72)$ &$\pos0.08412(80)$ &$\pos0.18243(91)$ \\
$0.80$ & $\pos0.27674(77)$ &$\pos0.12920(85)$ &$\pos0.1862(10)$ \\
$0.85$ & $\pos0.3179(13)$ &$\pos0.2057(13)$ &$\pos0.1918(12)$ \\
$0.90$ & $\pos0.4244(15)$ &$\pos0.3604(16)$ &$\pos0.1986(15)$ \\
$0.95$ & $\pos0.6492(19)$ &$\pos0.6615(19)$ &$\pos0.2186(23)$ \\
\hline\hline
\end{tabular}
\\[4ex]
\begin{tabular}{p{1cm}p{2.5cm}p{2.5cm}p{2.5cm}}
\hline\hline&&&\\[-3ex]
$v$ & \hspace{0.8cm}$c_-^{(V)0}$ &\hspace{0.8cm}$c_-^{(V)\parallel}$ &
       \hspace{0.8cm}$c_-^{(V)\perp}$\\&&&\\[-3ex]\hline
$0.10$ & $-0.1533(22)$ &$\pos0.3227(22)$ &$\pos0.1301(17)$ \\
$0.20$ & $-0.1547(13)$ &$\pos0.3220(13)$ &$\pos0.1274(10)$ \\
$0.30$ & $-0.1598(10)$ &$\pos0.3183(10)$ &$\pos0.12224(81)$ \\
$0.40$ & $-0.16838(86)$ &$\pos0.31106(85)$ &$\pos0.11325(72)$ \\
$0.50$ & $-0.18137(81)$ &$\pos0.29887(80)$ &$\pos0.09967(67)$ \\
$0.60$ & $-0.19460(80)$ &$\pos0.28291(80)$ &$\pos0.08532(68)$ \\
$0.70$ & $-0.21033(81)$ &$\pos0.25390(80)$ &$\pos0.06649(69)$ \\
$0.75$ & $-0.21513(84)$ &$\pos0.23150(83)$ &$\pos0.05808(73)$ \\
$0.80$ & $-0.21139(91)$ &$\pos0.20128(91)$ &$\pos0.05445(80)$ \\
$0.85$ & $-0.1931(11)$ &$\pos0.1530(12)$ &$\pos0.0561(10)$ \\
$0.90$ & $-0.1353(13)$ &$\pos0.0713(14)$ &$\pos0.0777(12)$ \\
$0.95$ & $\pos0.0615(23)$ &$-0.0832(23)$ &$\pos0.1347(22)$ \\
\hline\hline
\end{tabular}
\caption{Vector current matching coefficients for heavy quark mass $m=2.8$. The \asqtad\ action is 
used to discretize the light quark. The Lorentz indices can be 
timelike ($0$), parallel ($\parallel$) or perpendicular ($\perp$) to 
the frame velocity. The table shows the central value and error from the \vegas\ integration.} \label{tab:bsgamma_matching:c_match_full_vector}
\end{center}
\end{table*}

\begin{table*}
\renewcommand{\arraystretch}{1.3}
\begin{center}
\providecommand{\pos}{{\color{white}{-}}}
\begin{tabular}{p{1cm}p{2.5cm}p{2.5cm}p{2.5cm}p{2.5cm}}
\hline\hline&&&&\\[-3ex]
$v$ &\hspace{0.8cm}$c^{(T)0,1}$ &\hspace{0.8cm}$c^{(T)0,2}$ &\hspace{0.8cm}$c^{(T)1,2}$ &
     \hspace{0.8cm}$c^{(T)2,3}$\\&&&&\\[-3ex]\hline
$0.00$ & $\pos0.0762(12)$ &$\pos0.0761(12)$ &$\pos0.0763(12)$ &$\pos0.0761(12)$\\
\hline\hline
\end{tabular}
\\[4ex]
\begin{tabular}{p{1cm}p{2.5cm}p{2.5cm}p{2.5cm}p{2.5cm}}
\hline\hline&&&&\\[-3ex]
$v$ &\hspace{0.8cm} $c_+^{(T)0,\parallel}$ &\hspace{0.8cm}$c_+^{(T)0,\perp}$ 
      &\hspace{0.8cm}$c_+^{(T)\parallel,\perp}$ 
      &\hspace{0.8cm}$c_+^{(T)\perp,\perp}$\\&&&&\\[-3ex]\hline
$0.10$ & $-0.0507(24)$ &$-0.0159(21)$ &$-0.0157(22)$ &$-0.0463(24)$\\
$0.20$ & $-0.0559(17)$ &$-0.0206(16)$ &$-0.0214(16)$ &$-0.0526(17)$\\
$0.30$ & $-0.0675(16)$ &$-0.0325(15)$ &$-0.0321(15)$ &$-0.0652(16)$\\
$0.40$ & $-0.0861(15)$ &$-0.0496(15)$ &$-0.0497(15)$ &$-0.0846(15)$\\
$0.50$ & $-0.1148(16)$ &$-0.0748(15)$ &$-0.0750(15)$ &$-0.1136(16)$\\
$0.60$ & $-0.1605(17)$ &$-0.1135(16)$ &$-0.1136(16)$ &$-0.1595(17)$\\
$0.70$ & $-0.2302(19)$ &$-0.1666(18)$ &$-0.1668(18)$ &$-0.2295(19)$\\
$0.75$ & $-0.2791(21)$ &$-0.1984(21)$ &$-0.1992(21)$ &$-0.2782(21)$\\
$0.80$ & $-0.3379(26)$ &$-0.2280(25)$ &$-0.2292(25)$ &$-0.3368(26)$\\
$0.85$ & $-0.4006(34)$ &$-0.2397(35)$ &$-0.2397(35)$ &$-0.3990(34)$\\
$0.90$ & $-0.4463(53)$ &$-0.1688(53)$ &$-0.1696(53)$ &$-0.4441(53)$\\
$0.95$ & $-0.279(14)$ &$\pos0.224(14)$ &$\pos0.223(14)$ &$-0.279(14)$\\
\hline\hline
\end{tabular}
\\[4ex]
\begin{tabular}{p{1cm}p{2.5cm}p{2.5cm}p{2.5cm}p{2.5cm}}
\hline\hline&&&&\\[-3ex]
$v$ &\hspace{0.8cm} $c_-^{(T)0,\parallel}$ &\hspace{0.8cm}$c_-^{(T)0,\perp}$ 
    &\hspace{0.8cm}$c_-^{(T)\parallel,\perp}$ 
    &\hspace{0.8cm}$c_-^{(T)\perp,\perp}$\\&&&&\\[-3ex]\hline
$0.10$ & $\pos0.1184(20)$ &$\pos0.0849(18)$ &$\pos0.0871(18)$ &$\pos0.1189(20)$\\
$0.20$ & $\pos0.1147(11)$ &$\pos0.0839(11)$ &$\pos0.0855(11)$ &$\pos0.1149(11)$\\
$0.30$ & $\pos0.10781(85)$ &$\pos0.07967(86)$ &$\pos0.08054(86)$ &$\pos0.10774(85)$\\
$0.40$ & $\pos0.09607(74)$ &$\pos0.07144(76)$ &$\pos0.07214(77)$ &$\pos0.09583(74)$\\
$0.50$ & $\pos0.07857(69)$ &$\pos0.05896(72)$ &$\pos0.05912(72)$ &$\pos0.07848(69)$\\
$0.60$ & $\pos0.05889(69)$ &$\pos0.04414(72)$ &$\pos0.04412(72)$ &$\pos0.05860(69)$\\
$0.70$ & $\pos0.03355(70)$ &$\pos0.02162(73)$ &$\pos0.02142(73)$ &$\pos0.03332(70)$\\
$0.75$ & $\pos0.02163(74)$ &$\pos0.00812(77)$ &$\pos0.00787(76)$ &$\pos0.02131(74)$\\
$0.80$ & $\pos0.01460(82)$ &$-0.00494(84)$ &$-0.00522(84)$ &$\pos0.01425(82)$\\
$0.85$ & $\pos0.0146(10)$ &$-0.0199(10)$ &$-0.0205(10)$ &$\pos0.0142(10)$\\
$0.90$ & $\pos0.0361(13)$ &$-0.0310(13)$ &$-0.0312(13)$ &$\pos0.0357(13)$\\
$0.95$ & $\pos0.0980(22)$ &$-0.0125(22)$ &$-0.0114(22)$ &$\pos0.0976(22)$\\
\hline\hline
\end{tabular}

\caption{Tensor current matching coefficients for heavy quark mass $m=2.8$. The \asqtad\ action is 
used to discretize the light quark.} \label{tab:bsgamma_matching:c_match_full}
\end{center}
\end{table*}

\begin{table*}
\renewcommand{\arraystretch}{1.3}
\begin{center}
\providecommand{\pos}{{\color{white}{-}}}
\begin{tabular}{p{1cm}p{2.5cm}p{2.5cm}p{2.5cm}}
\hline\hline&&&\\[-3ex]
$v$ &\hspace{0.8cm}$c^{(V)0}$ &\hspace{0.8cm}$c^{(V)1}$ &\hspace{0.8cm}$c^{(V)2}$\\&&&\\[-3ex]\hline
$0.00$ & $-0.10173(91)$ &$-0.03811(87)$ &$-0.03825(87)$\\
\hline\hline
\end{tabular}
\\[4ex]
\begin{tabular}{p{1cm}p{2.5cm}p{2.5cm}p{2.5cm}}
\hline&&&\\[-3ex]
$v$ & \hspace{0.8cm}$c_+^{(V)0}$ &\hspace{0.8cm}$c_+^{(V)\parallel}$ 
 &\hspace{0.8cm}$c_+^{(V)\perp}$\\&&&\\[-3ex]\hline\hline
$0.10$ & $-0.1323(22)$ &$-0.1980(22)$ &$-0.1593(20)$ \\
$0.20$ & $-0.1318(15)$ &$-0.1942(15)$ &$-0.1575(14)$ \\
$0.30$ & $-0.1286(13)$ &$-0.1902(13)$ &$-0.1513(12)$ \\
$0.40$ & $-0.1226(12)$ &$-0.1835(12)$ &$-0.1417(12)$ \\
$0.50$ & $-0.1147(11)$ &$-0.1732(12)$ &$-0.1276(11)$ \\
$0.60$ & $-0.1079(11)$ &$-0.1636(11)$ &$-0.1116(11)$ \\
$0.70$ & $-0.0982(11)$ &$-0.1454(11)$ &$-0.0895(11)$ \\
$0.75$ & $-0.0906(11)$ &$-0.1307(11)$ &$-0.0760(12)$ \\
$0.80$ & $-0.0775(11)$ &$-0.1067(11)$ &$-0.0610(12)$ \\
$0.85$ & $-0.0465(12)$ &$-0.0578(12)$ &$-0.0412(14)$ \\
$0.90$ & $\pos0.0268(13)$ &$\pos0.0411(14)$ &$-0.0157(16)$ \\
$0.95$ & $\pos0.2011(22)$ &$\pos0.2624(21)$ &$\pos0.0229(23)$ \\
\hline\hline
\end{tabular}
\\[4ex]
\begin{tabular}{p{1cm}p{2.5cm}p{2.5cm}p{2.5cm}}
\hline\hline&&&\\[-3ex]
$v$ & \hspace{0.8cm}$c_-^{(V)0}$ &\hspace{0.8cm}$c_-^{(V)\parallel}$ 
 &\hspace{0.8cm}$c_-^{(V)\perp}$\\&&&\\[-3ex]\hline
$0.10$ & $\pos0.0322(21)$ &$\pos0.1574(20)$ &$\pos0.1218(16)$ \\
$0.20$ & $\pos0.0303(13)$ &$\pos0.1562(12)$ &$\pos0.1183(10)$ \\
$0.30$ & $\pos0.0269(10)$ &$\pos0.15322(93)$ &$\pos0.11199(78)$ \\
$0.40$ & $\pos0.02062(84)$ &$\pos0.14692(82)$ &$\pos0.10099(70)$ \\
$0.50$ & $\pos0.01122(78)$ &$\pos0.13632(76)$ &$\pos0.08488(66)$ \\
$0.60$ & $\pos0.00253(77)$ &$\pos0.12358(76)$ &$\pos0.06641(67)$ \\
$0.70$ & $-0.00754(78)$ &$\pos0.10023(77)$ &$\pos0.04171(68)$ \\
$0.75$ & $-0.00961(81)$ &$\pos0.08269(80)$ &$\pos0.02950(72)$ \\
$0.80$ & $-0.00407(88)$ &$\pos0.05947(88)$ &$\pos0.02188(79)$ \\
$0.85$ & $\pos0.0132(10)$ &$\pos0.0209(10)$ &$\pos0.01954(93)$ \\
$0.90$ & $\pos0.0648(13)$ &$-0.0436(13)$ &$\pos0.0357(12)$ \\
$0.95$ & $\pos0.2373(22)$ &$-0.1771(23)$ &$\pos0.0933(22)$ \\
\hline\hline
\end{tabular}
\caption{Vector current matching coefficients for heavy quark mass $m=2.8$. The \hisq\ action is 
used to discretize the light quark.} 
\label{tab:bsgamma_matching:c_match_full_hisq_vector}
\end{center}
\end{table*}

\begin{table*}
\renewcommand{\arraystretch}{1.3}
\begin{center}
\providecommand{\pos}{{\color{white}{-}}}
\begin{tabular}{p{1cm}p{2.5cm}p{2.5cm}p{2.5cm}p{2.5cm}}
\hline\hline&&&&\\[-3ex]
$v$ &\hspace{0.8cm}$c^{(T)0,1}$ &\hspace{0.8cm}$c^{(T)0,2}$ 
&\hspace{0.8cm}$c^{(T)1,2}$ &\hspace{0.8cm}$c^{(T)2,3}$\\&&&&\\[-3ex]\hline
$0.00$ & $-0.2317(15)$ &$-0.2318(15)$ &$-0.2316(15)$ &$-0.2317(15)$\\
\hline\hline
\end{tabular}
\\[4ex]
\begin{tabular}{p{1cm}p{2.5cm}p{2.5cm}p{2.5cm}p{2.5cm}}
\hline\hline&&&&\\[-3ex]
$v$ &\hspace{0.8cm} $c_+^{(T)0,\parallel}$ &\hspace{0.8cm}$c_+^{(T)0,\perp}$ 
&\hspace{0.8cm}$c_+^{(T)\parallel,\perp}$ &\hspace{0.8cm}$c_+^{(T)\perp,\perp}$\\&&&&\\[-3ex]\hline
$0.10$ & $-0.3610(25)$ &$-0.3330(23)$ &$-0.3305(23)$ &$-0.3555(25)$\\
$0.20$ & $-0.3643(19)$ &$-0.3380(18)$ &$-0.3375(18)$ &$-0.3614(19)$\\
$0.30$ & $-0.3741(17)$ &$-0.3500(17)$ &$-0.3496(17)$ &$-0.3726(17)$\\
$0.40$ & $-0.3890(17)$ &$-0.3691(17)$ &$-0.3690(17)$ &$-0.3881(17)$\\
$0.50$ & $-0.4120(17)$ &$-0.3974(17)$ &$-0.3970(17)$ &$-0.4109(17)$\\
$0.60$ & $-0.4480(18)$ &$-0.4410(18)$ &$-0.4414(18)$ &$-0.4469(18)$\\
$0.70$ & $-0.5022(21)$ &$-0.5040(20)$ &$-0.5043(20)$ &$-0.5011(21)$\\
$0.75$ & $-0.5394(23)$ &$-0.5437(22)$ &$-0.5441(22)$ &$-0.5384(23)$\\
$0.80$ & $-0.5829(27)$ &$-0.5843(26)$ &$-0.5854(26)$ &$-0.5827(27)$\\
$0.85$ & $-0.6264(35)$ &$-0.6102(34)$ &$-0.6103(34)$ &$-0.6264(35)$\\
$0.90$ & $-0.6497(53)$ &$-0.5754(53)$ &$-0.5761(53)$ &$-0.6490(53)$\\
$0.95$ & $-0.458(14)$ &$-0.240(14)$ &$-0.240(14)$ &$-0.459(14)$\\
\hline\hline
\end{tabular}
\\[4ex]
\begin{tabular}{p{1cm}p{2.5cm}p{2.5cm}p{2.5cm}p{2.5cm}}
\hline\hline&&&&\\[-3ex]
$v$ &\hspace{0.8cm} $c_-^{(T)0,\parallel}$ &\hspace{0.8cm}$c_-^{(T)0,\perp}$ 
&\hspace{0.8cm}$c_-^{(T)\parallel,\perp}$ &\hspace{0.8cm}$c_-^{(T)\perp,\perp}$\\&&&&\\[-3ex]\hline
$0.10$ & $\pos0.1230(19)$ &$\pos0.0964(17)$ &$\pos0.0955(18)$ &$\pos0.1234(19)$\\
$0.20$ & $\pos0.1187(11)$ &$\pos0.0939(11)$ &$\pos0.0938(11)$ &$\pos0.1187(11)$\\
$0.30$ & $\pos0.11138(82)$ &$\pos0.09018(84)$ &$\pos0.09064(84)$ &$\pos0.11126(82)$\\
$0.40$ & $\pos0.09914(72)$ &$\pos0.08398(75)$ &$\pos0.08432(75)$ &$\pos0.09898(72)$\\
$0.50$ & $\pos0.08099(67)$ &$\pos0.07380(70)$ &$\pos0.07417(70)$ &$\pos0.08075(67)$\\
$0.60$ & $\pos0.05968(67)$ &$\pos0.06302(70)$ &$\pos0.06342(70)$ &$\pos0.05937(67)$\\
$0.70$ & $\pos0.03107(68)$ &$\pos0.04637(71)$ &$\pos0.04668(71)$ &$\pos0.03060(68)$\\
$0.75$ & $\pos0.01635(73)$ &$\pos0.03647(75)$ &$\pos0.03669(75)$ &$\pos0.01583(72)$\\
$0.80$ & $\pos0.00591(80)$ &$\pos0.02745(82)$ &$\pos0.02778(82)$ &$\pos0.00529(80)$\\
$0.85$ & $\pos0.00098(94)$ &$\pos0.0168(10)$ &$\pos0.0169(10)$ &$\pos0.00026(94)$\\
$0.90$ & $\pos0.0146(12)$ &$\pos0.0102(12)$ &$\pos0.0108(12)$ &$\pos0.0141(12)$\\
$0.95$ & $\pos0.0705(22)$ &$\pos0.0299(22)$ &$\pos0.0302(22)$ &$\pos0.0696(22)$\\
\hline\hline
\end{tabular}
\caption{Tensor current matching coefficients for heavy quark mass $m=2.8$. The \hisq\ action is 
used to discretize the light quark.} \label{tab:bsgamma_matching:c_match_full_HISQ}
\end{center}
\end{table*}


\begin{table*}
\renewcommand{\arraystretch}{1.3}
\begin{center}
\providecommand{\pos}{{\color{white}{-}}}
\providecommand{\pos}{{\color{white}{-}}}
\begin{tabular}{p{1cm}p{2.5cm}p{2.5cm}p{2.5cm}}
\hline\hline&&&\\[-3ex]
$v$ &\hspace{0.8cm}$c^{(V)0}$ &\hspace{0.8cm}$c^{(V)1}$ &\hspace{0.8cm}$c^{(V)2}$\\&&&\\[-3ex]\hline
$0.00$ & $-0.06534(59)$ &$\pos0.33741(41)$ &$\pos0.33681(41)$\\
\hline\hline
\end{tabular}
\\[4ex]
\begin{tabular}{p{1cm}p{2.5cm}p{2.5cm}p{2.5cm}}
\hline\hline&&&\\[-3ex]
$v$ & \hspace{0.8cm}$c_+^{(V)0}$ &\hspace{0.8cm}$c_+^{(V)\parallel}$ &\hspace{0.8cm}$c_+^{(V)\perp}$\\&&&\\[-3ex]\hline
$0.01$ & $\pos0.246(18)$ &$-0.185(18)$ &$\pos0.162(14)$ \\
$0.10$ & $\pos0.2164(21)$ &$-0.1954(22)$ &$\pos0.1514(18)$ \\
$0.20$ & $\pos0.2147(13)$ &$-0.1924(14)$ &$\pos0.1501(12)$ \\
$0.30$ & $\pos0.2170(10)$ &$-0.1849(12)$ &$\pos0.1506(10)$ \\
\hline\hline
\end{tabular}
\\[4ex]
\begin{tabular}{p{1cm}p{2.5cm}p{2.5cm}p{2.5cm}}
\hline\hline&&&\\[-3ex]
$v$ & \hspace{0.8cm}$c_-^{(V)0}$ &\hspace{0.8cm}$c_-^{(V)\parallel}$ &\hspace{0.8cm}$c_-^{(V)\perp}$\\&&&\\[-3ex]\hline
$0.01$ & $-0.294(18)$ &$\pos0.530(18)$ &$\pos0.185(14)$ \\
$0.10$ & $-0.2831(21)$ &$\pos0.5337(22)$ &$\pos0.1848(17)$ \\
$0.20$ & $-0.2829(13)$ &$\pos0.5353(13)$ &$\pos0.1847(11)$ \\
$0.30$ & $-0.2889(11)$ &$\pos0.5333(11)$ &$\pos0.18050(89)$ \\
\hline\hline
\end{tabular}
\caption{Vector current matching coefficients for heavy quark mass $m=1.9$. The \asqtad\ action is 
used to discretize the light quark.} \label{tab:bsgamma_matching:c_match_full_vector_m=1.9}
\end{center}
\end{table*}

\begin{table*}
\renewcommand{\arraystretch}{1.3}
\begin{center}
\providecommand{\pos}{{\color{white}{-}}}
\providecommand{\pos}{{\color{white}{-}}}
\begin{tabular}{p{1cm}p{2.5cm}p{2.5cm}p{2.5cm}p{2.5cm}}
\hline\hline&&&&\\[-3ex]
$v$ &\hspace{0.8cm}$c^{(T)0,1}$ &\hspace{0.8cm}$c^{(T)0,2}$ &\hspace{0.8cm}$c^{(T)1,2}$ &\hspace{0.8cm}$c^{(T)2,3}$\\&&&&\\[-3ex]\hline
$0.00$ & $\pos0.3410(16)$ &$\pos0.3411(16)$ &$\pos0.3414(16)$ &$\pos0.3412(16)$\\
\hline\hline
\end{tabular}
\\[4ex]
\begin{tabular}{p{1cm}p{2.5cm}p{2.5cm}p{2.5cm}p{2.5cm}}
\hline\hline&&&&\\[-3ex]
$v$ &\hspace{0.8cm} $c_+^{(T)0,\parallel}$ &\hspace{0.8cm}$c_+^{(T)0,\perp}$ &\hspace{0.8cm}$c_+^{(T)\parallel,\perp}$ &\hspace{0.8cm}$c_+^{(T)\perp,\perp}$\\&&&&\\[-3ex]\hline
$0.01$ & $\pos0.141(18)$ &$\pos0.210(14)$ &$\pos0.205(14)$ &$\pos0.167(18)$\\
$0.10$ & $\pos0.1524(26)$ &$\pos0.2094(24)$ &$\pos0.2085(24)$ &$\pos0.1577(26)$\\
$0.20$ & $\pos0.1448(21)$ &$\pos0.2017(20)$ &$\pos0.2017(20)$ &$\pos0.1476(21)$\\
$0.30$ & $\pos0.1319(19)$ &$\pos0.1907(18)$ &$\pos0.1903(18)$ &$\pos0.1339(19)$\\
\hline\hline
\end{tabular}
\\[4ex]
\begin{tabular}{p{1cm}p{2.5cm}p{2.5cm}p{2.5cm}p{2.5cm}}
\hline\hline&&&&\\[-3ex]
$v$ &\hspace{0.8cm} $c_-^{(T)0,\parallel}$ &\hspace{0.8cm}$c_-^{(T)0,\perp}$ &\hspace{0.8cm}$c_-^{(T)\parallel,\perp}$ &\hspace{0.8cm}$c_-^{(T)\perp,\perp}$\\&&&&\\[-3ex]\hline
$0.01$ & $\pos0.162(18)$ &$\pos0.119(14)$ &$\pos0.119(14)$ &$\pos0.190(18)$\\
$0.10$ & $\pos0.1784(20)$ &$\pos0.1268(18)$ &$\pos0.1285(18)$ &$\pos0.1814(20)$\\
$0.20$ & $\pos0.1777(12)$ &$\pos0.1267(12)$ &$\pos0.1273(12)$ &$\pos0.1785(12)$\\
$0.30$ & $\pos0.1712(10)$ &$\pos0.12228(94)$ &$\pos0.12256(94)$ &$\pos0.17183(95)$\\
\hline\hline
\end{tabular}
\caption{Tensor current matching coefficients for heavy quark mass $m=1.9$. The \asqtad\ action is 
used to discretize the light quark.} \label{tab:bsgamma_matching:c_match_full_m=1.9}
\end{center}
\end{table*}

\begin{table*}
\renewcommand{\arraystretch}{1.3}
\begin{center}
\providecommand{\pos}{{\color{white}{-}}}
\providecommand{\pos}{{\color{white}{-}}}
\begin{tabular}{p{1cm}p{2.5cm}p{2.5cm}p{2.5cm}}
\hline\hline&&&\\[-3ex]
$v$ &\hspace{0.8cm}$c^{(V)0}$ &\hspace{0.8cm}$c^{(V)1}$ &\hspace{0.8cm}$c^{(V)2}$\\&&&\\[-3ex]\hline
$0.00$ & $-0.15556(94)$ &$\pos0.01365(88)$ &$\pos0.01296(88)$\\
\hline\hline
\end{tabular}
\\[4ex]
\begin{tabular}{p{1cm}p{2.5cm}p{2.5cm}p{2.5cm}}
\hline\hline&&&\\[-3ex]
$v$ & \hspace{0.8cm}$c_+^{(V)0}$ &\hspace{0.8cm}$c_+^{(V)\parallel}$ &\hspace{0.8cm}$c_+^{(V)\perp}$\\&&&\\[-3ex]\hline
$0.01$ & $-0.095(17)$ &$-0.273(17)$ &$-0.164(13)$ \\
$0.10$ & $-0.1180(23)$ &$-0.2782(22)$ &$-0.1721(20)$ \\
$0.20$ & $-0.1198(16)$ &$-0.2779(16)$ &$-0.1720(15)$ \\
$0.30$ & $-0.1176(13)$ &$-0.2744(14)$ &$-0.1689(13)$ \\
\hline\hline
\end{tabular}
\\[4ex]
\begin{tabular}{p{1cm}p{2.5cm}p{2.5cm}p{2.5cm}}
\hline\hline&&&\\[-3ex]
$v$ & \hspace{0.8cm}$c_-^{(V)0}$ &\hspace{0.8cm}$c_-^{(V)\parallel}$ &\hspace{0.8cm}$c_-^{(V)\perp}$\\&&&\\[-3ex]\hline
$0.01$ & $-0.055(17)$ &$\pos0.283(17)$ &$\pos0.189(13)$ \\
$0.10$ & $-0.0359(21)$ &$\pos0.2908(21)$ &$\pos0.1840(17)$ \\
$0.20$ & $-0.0366(13)$ &$\pos0.2926(13)$ &$\pos0.1831(11)$ \\
$0.30$ & $-0.0415(10)$ &$\pos0.2914(10)$ &$\pos0.17766(88)$ \\
\hline\hline
\end{tabular}
\caption{Vector current matching coefficients for heavy quark mass $m=1.9$. The \hisq\ action is 
used to discretize the light quark.} 
\label{tab:bsgamma_matching:c_match_full_hisq_vector_m=1.9}
\end{center}
\end{table*}

\begin{table*}
\renewcommand{\arraystretch}{1.3}
\begin{center}
\providecommand{\pos}{{\color{white}{-}}}
\begin{tabular}{p{1cm}p{2.5cm}p{2.5cm}p{2.5cm}p{2.5cm}}
\hline\hline&&&&\\[-3ex]
$v$ &\hspace{0.8cm}$c^{(T)0,1}$ &\hspace{0.8cm}$c^{(T)0,2}$ &\hspace{0.8cm}$c^{(T)1,2}$ &\hspace{0.8cm}$c^{(T)2,3}$\\&&&&\\[-3ex]\hline
$0.00$ & $\pos0.0171(17)$ &$\pos0.0175(17)$ &$\pos0.0172(17)$ &$\pos0.0173(17)$\\
\hline\hline
\end{tabular}
\\[4ex]
\begin{tabular}{p{1cm}p{2.5cm}p{2.5cm}p{2.5cm}p{2.5cm}}
\hline\hline&&&&\\[-3ex]
$v$ &\hspace{0.8cm} $c_+^{(T)0,\parallel}$ &\hspace{0.8cm}$c_+^{(T)0,\perp}$ &\hspace{0.8cm}$c_+^{(T)\parallel,\perp}$ &\hspace{0.8cm}$c_+^{(T)\perp,\perp}$\\&&&&\\[-3ex]\hline
$0.01$ & $-0.199(17)$ &$-0.123(13)$ &$-0.112(13)$ &$-0.174(17)$\\
$0.10$ & $-0.1800(27)$ &$-0.1163(25)$ &$-0.1150(25)$ &$-0.1763(27)$\\
$0.20$ & $-0.1863(22)$ &$-0.1231(21)$ &$-0.1229(21)$ &$-0.1840(22)$\\
$0.30$ & $-0.1966(20)$ &$-0.1347(20)$ &$-0.1345(20)$ &$-0.1950(20)$\\
\hline\hline
\end{tabular}
\\[4ex]
\begin{tabular}{p{1cm}p{2.5cm}p{2.5cm}p{2.5cm}p{2.5cm}}
\hline\hline&&&&\\[-3ex]
$v$ &\hspace{0.8cm} $c_-^{(T)0,\parallel}$ &\hspace{0.8cm}$c_-^{(T)0,\perp}$ &\hspace{0.8cm}$c_-^{(T)\parallel,\perp}$ &\hspace{0.8cm}$c_-^{(T)\perp,\perp}$\\&&&&\\[-3ex]\hline
$0.01$ & $\pos0.208(17)$ &$\pos0.135(13)$ &$\pos0.125(14)$ &$\pos0.183(17)$\\
$0.10$ & $\pos0.1915(19)$ &$\pos0.1285(17)$ &$\pos0.1287(18)$ &$\pos0.1889(19)$\\
$0.20$ & $\pos0.1889(12)$ &$\pos0.1282(12)$ &$\pos0.1287(12)$ &$\pos0.1876(12)$\\
$0.30$ & $\pos0.18199(93)$ &$\pos0.12518(94)$ &$\pos0.12555(94)$ &$\pos0.18114(93)$\\
\hline\hline
\end{tabular}
\caption{Tensor current matching coefficients for heavy quark mass $am=1.9$. The \hisq\ action is 
used to discretize the light quark.} \label{tab:bsgamma_matching:c_match_full_HISQ_m=1.9}
\end{center}
\end{table*}
\clearpage

\bibliographystyle{apsrev4-1}
\bibliography{hpqcd_mnrqcd_currentmatching}

\end{document}